\documentclass[aps,prd,nofootinbib,twocolumn,groupedaddress]{revtex4}
\usepackage{dcolumn}
\usepackage{bm}
\usepackage{pdfpages}
\usepackage{amsmath}    
\usepackage{amsfonts} 
\usepackage{amssymb}
\usepackage{mathptmx}      
\usepackage{graphicx}   
\usepackage{mathtools}
\usepackage{graphics}
\usepackage{epsfig}
\usepackage{tabularx}   
\usepackage{multirow}
\usepackage{bibentry}
\usepackage{braket}
\usepackage{algorithm}
\usepackage{algorithmic}
\usepackage{listings}
\usepackage{setspace}
\usepackage{fancyhdr}
\usepackage{slashed}
\usepackage{color}
\usepackage{xfrac}
\usepackage[pdftex, pdfborder= 0 0 0, citecolor=blue, urlcolor=blue, linkcolor=blue, colorlinks=true, bookmarksopen=true]{hyperref}

\def\la{\langle}
\def\ra{\rangle}

\def\be{\begin{equation}}
\def\ee{\end{equation}}
\def\bea{\begin{eqnarray}}
\def\eea{\end{eqnarray}}
\begin{document}
\title{Pion off-shell electromagnetic form factors: Data extraction and model analysis}
\author{ Ho-Meoyng Choi}
\affiliation{ Department of Physics, Teachers College, Kyungpook National University,
     Daegu 41566, Korea}
          
\author{  T. Frederico}
\affiliation{ Instituto Tecnol\'ogico de
Aeron\'autica, 12.228-900 S\~ao Jos\'e dos Campos, SP,
Brazil}

 \author{Chueng-Ryong Ji}
\affiliation{ Department of Physics, North Carolina State University,
Raleigh, NC 27695-8202, USA} 

\author{  J. P. B. C. de Melo}
\affiliation{ Laborat\'orio de F\'\i sica Te\'orica e Computacional, Universidade Cruzeiro do Sul/Universidade Cidade
de S\~ao Paulo, 01506-000, S\~ao Paulo SP, Brazil
}

\begin{abstract}
 We investigate the pion electromagnetic half-off-shell
form factors, which parametrize the matrix element of the charged
pion electromagnetic current with one leg off-mass-shell
and the other leg on-mass-shell, using an exactly solvable manifestly
covariant model of a $(3+1)$ dimensional fermion field theory.
The model provides a three-dimensional imaging of the two off-shell pion form factors
$F_1$ and $F_2$ as a function of $(Q^2,t)$, which
are related to each other satisfying the Ward-Takahashi identity.
The normalization of the renormalized charge form factor $F_1$ is fixed by $F_1(Q^2=0, t=m^2_\pi)=1$ while the other form factor $F_2$ vanishes, i.e. 
$F_2(Q^2, t=m^2_\pi)=0$ for any value of $Q^2$ due to the time-reversal invariance of the strong interaction. 
We define the new form factor $g(Q^2,t)=F_2(Q^2,t)/(t- m^2_\pi)$ and
find that $g(Q^2,t)$ can be measurable in the on-mass-shell limit. In particular, $g(Q^2=0, t=m^2_\pi)$
is related with the pion charge radius.
We also compare our form factors  with the data extracted from the pion electroproduction reaction for both the off-shell region ($t<0$) 
and the on-shell limit ($t \rightarrow m_\pi^2$).
\end{abstract}
\maketitle

\section{Introduction}
Electromagnetic (EM) form factors of  hadrons are the important physical observables providing the EM
information on the bound-state properties of hadrons and their internal structures of quarks and gluons.
The pion  is  the simplest hadronic system, the valence structure of which is 
a bound state of a quark and an antiquark, and is known to be parametrized by a single on-mass-shell (or simply on-shell) EM form factor,
$F_\pi(Q^2)$, which depends on the 4-momentum squared $q^2(=-Q^2)$ of the virtual photon.

\begin{figure}
\begin{center}
\includegraphics[height=3cm, width=4cm]{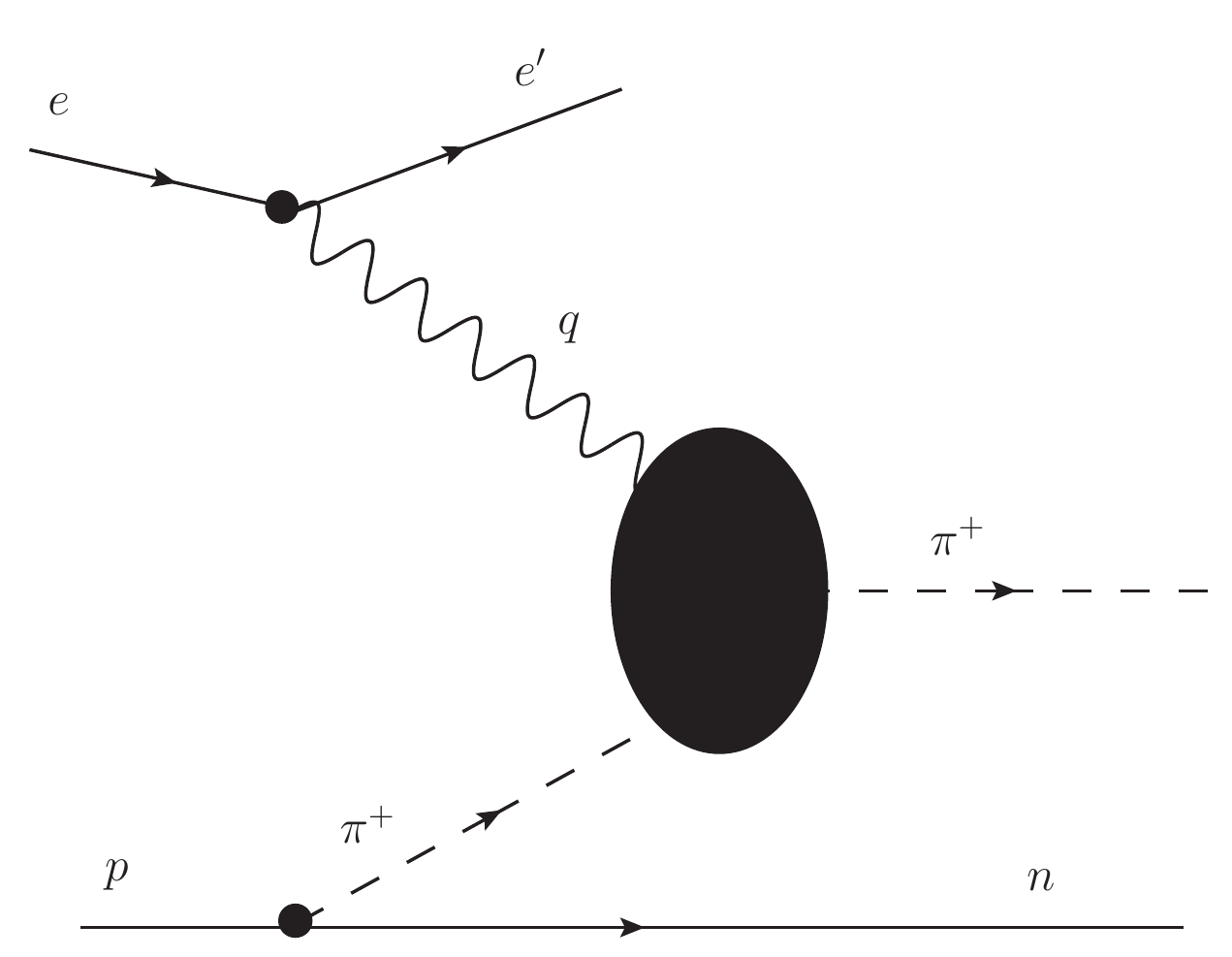}
\caption{\label{fig1} $ep\rightarrow e'\pi^+ n$ scattering.}
\end{center}
\end{figure}

The form factor $F_\pi(Q^2)$ for the
low spacelike momentum transfers ($Q^2 < 0.3$ GeV$^2$) has been measured directly by elastic scattering of high-energy
mesons off atomic electrons~\cite{Dally1,Dally2,Amen1,Amen2}. However, the extraction
of  $F_\pi(Q^2)$  to higher $Q^2$ regions through elastic scattering is very difficult experimentally
mainly due to the limitation of the availability of accelerators to produce high-energy and high-current beams 
of unstable particles and detectors for identifying and measuring the scattered particles at very forward angles~\cite{Marco}. 
Thus, $F_\pi(Q^2)$ for the higher $Q^2$ values
has been extracted from the pion electroproduction reaction by exploiting
the nucleon's pion cloud as a target, which may be regarded as the exclusive version
of the Sullivan process~\cite{Sull}.
That is, $F_\pi(Q^2)$  has been extracted from the measurements of the cross sections for the reaction $^1{\rm H}(e,e'\pi^+)n$ (see Fig.~\ref{fig1})
up to values of $Q^2=3.91$ GeV$^2$~\cite{Blok2008,Huber2008,JLab3,JLab4,JLab5}.
The longitudinal part of the cross section from the pion electroproduction  encodes the meson exchange process, 
in which the virtual photon couples to a virtual pion inside the nucleon. This process is expected to dominate at small values 
of the 4-momentum transfer $t(<0)$, allowing for the determination of the pion form factor. 

However, the main problem in using the electroproduction process as a tool for accessing 
a ``pion target" is that the pions in a nucleon's cloud are not real (on-shell) but virtual (off-shell) particles.
Accordingly, one cannot access the form factor at the exact pion pole in the actual experiment
as the extrapolation to $t \to m_\pi^2$ involves the disallowed kinematic region of the electroproduction
($t < 0$).
This may raise some questions about the validity of the  extrapolation from the off-shell  results to the on-shell
limit. Furthermore, the EM structure of 
the off-shell hadron is more complicated than the on-shell hadron and involves more form 
factors~\cite{Rudy,Weiss,Craig,CraigK,N1,N2,N3,N4,N5}.
For instance, the off-shell EM structure of the pseudoscalar meson~\cite{Rudy,Weiss} 
requires two form factors~\cite{Nis,Bar}, which are related by the Ward-Takahashi identity(WTI)~\cite{Ward,Taka}.
The off-shell electromagnetic form factors for the boson
bound state have been calculated in~\cite{Naus1998}
using the light-front field theory and the nonvanishing zero modes  were found to
be crucial to preserve the WTI.
While there have been some theoretical studies on the off-shell pion EM form 
factors using the chiral perturbation theory~\cite{Rudy}, Nambu-Jona-Lasinio model~\cite{Weiss},
and the continuum methods for the strong-interaction bound-state problem~\cite{Craig,CraigK},
a further systematic study on the off-shell form factors of the pion is still required.

In this work, we explore the electromagnetic off-shell effects for the 
pion using an exactly solvable manifestly covariant model of $(3+1)$-dimensional fermion field theory
and compare the two off-shell  form factors  $F_1(Q^2,t)$ and $F_2(Q^2,t)$
with the data extracted 
from the pion electroproduction reaction~\cite{Blok2008,Huber2008}.
The aim of this paper is to provide at least a clear example of demonstration discussing the validity of
the extrapolation of the off-shell results ($t \neq m_\pi^2$) to the on-shell limit ($t=m_\pi^2$) for the pion.
We exhibit $F_{1(2)}(Q^2,t)$
not only for the spacelike region ($Q^2>0$) but also  for the timelike
region ($Q^2<0$), providing the three-dimensional (3D) imaging of $F_1$ and $F_2$
in terms of $(Q^2, t)$ values. 

We organize this work as follows. In Sec. \ref{off-shellpionFF}, we review the formulation of 
$F_1$ and $F_2$ satisfying the WTI,
in which two form factors are necessary to define the off-shell matrix elements of the pion EM current. 
In addition, we provide a sum rule,
coined here as the master equation, which we obtain from the WTI that the form factors must obey 
regardless of whether they are on-shell or off-shell. 
While $F_2(Q^2,t)$ is zero as $t\to m^2_\pi$, 
we find a new measurable form factor in the on-shell limit by defining 
$g(Q^2,t)=F_2(Q^2,t)/(t- m^2_\pi)$.  
 Especially, we show that $g(Q^2=0, t=m^2_\pi)$ is found to be related with the pion charge radius.
In Sec.~\ref{pionhosffmodel}, we present the analytic covariant model calculation of 
$F_1$ and $F_2$
confirming that the model satisfies the master equation given by Eq.(\ref{eq7}) as well as the WTI. 
We also discuss the charge renormalization for $F_1(Q^2,t)$ 
together with the relation between the coupling $g_{\pi q{\bar q}}$ of the $\pi q\bar{q}$ vertex
and the pion decay constant $f_\pi$.
In Sec.~\ref{xsection}, we present 3D imaging of $F_1(Q^2,t), F_2(Q^2,t)$ and $g(Q^2,t)$ and 
compare them with the available data extracted from the pion
electroproduction reaction for both the off-shell region ($t<0$) and the on-shell limit ($t \rightarrow m_\pi^2$).
A summary of the main results follows in  Sec.~\ref{summary}. 
In the Appendix, the explicit derivation of Eqs.~(19) and ~(20) is briefly summarized.

\begin{figure}
\begin{center}
\includegraphics[height=3cm, width=4cm]{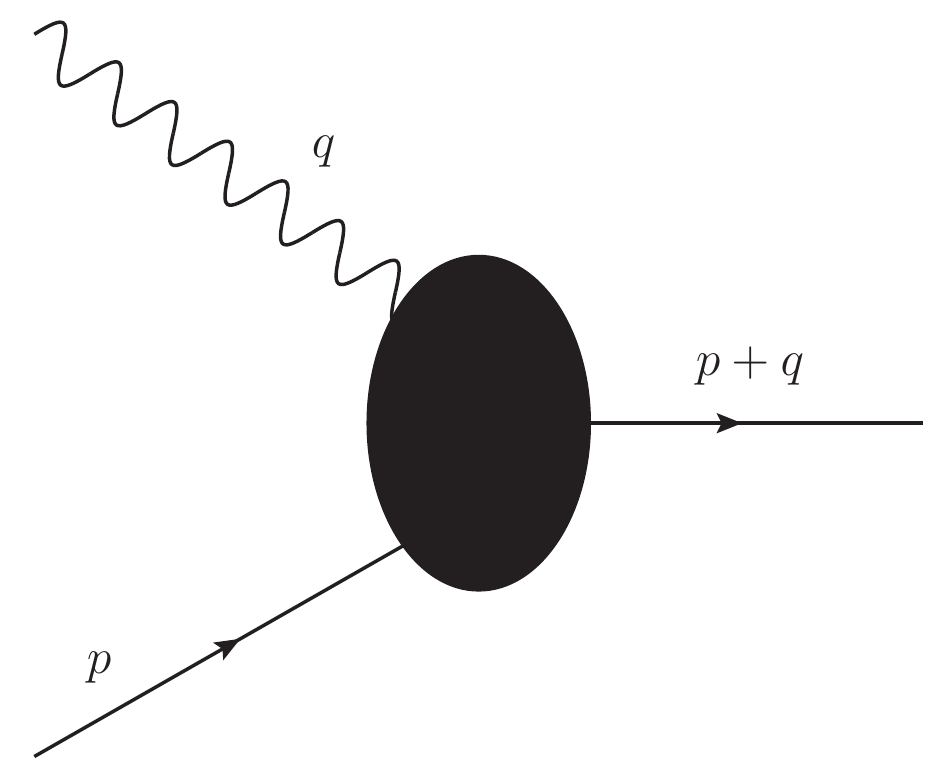}
\caption{\label{fig2}Electromagnetic charged pion scattering with the form factors depicted by the black blob.}
\end{center}
\end{figure}

\section{Off-shell Pion Electromagnetic Form Factors}  \label{off-shellpionFF}

Using the invariance of the strong interaction under charge conjugation, one finds that the electromagnetic form factors of antiparticles 
are just the negative of those of the particles. Therefore, the $\pi^0$ and $\eta$ do not have any electromagnetic form factors even for the off-mass-shell case. 
However, the charged pions allow the electromagnetic form factors depicted in Fig.~\ref{fig2}. 
The most general parametrization of the vertex function $\Gamma^\mu$ for the off-shell electromagnetic form factors of the charged pion  is given 
in terms of the initial and final 4-momenta, $p^\mu$ and ~$p'^{\mu}$, as~\cite{Rudy}
\begin{equation}\label{eq1}
\Gamma_\mu(p, p') =~ (p'+p)_\mu ~G_1(q^2,p^2,p'^{2}) + q_\mu ~G_2(q^2,p^2,p'^{2})~,
\end{equation}
where  $q=p'-p$ is the 4-momentum transfer of the virtual photon at the vertex. 
This off-shell vertex satisfies the WTI~\cite{Rudy}
\begin{equation}\label{eq2}
 q^\mu \Gamma_\mu (p,p') =~\Delta^{-1} (p') -  \Delta^{-1}(p),
 \end{equation}  
 where
 \begin{equation}\label{eq3}
 \Delta (p)~=~ \frac{1}{p^2 - m^2_\pi -\Pi(p^2) + \imath \epsilon} ,
 \end{equation}
is the full renormalized propagator~\cite{Rudy} and
the renormalized pion self-energy $\Pi(p^2)$ is constrained by the on-mass-shell condition $\Pi(m^2_\pi)=0$.

From the WTI given by Eq.~(\ref{eq2}),  we get the following constraint on the off-shell form factors 
$G_1$ and $G_2$:
\bea\label{WTI1}
&&(p'^2 - p^2) G_1 (q^2,p^2, p'^2)  + q^2 G_2 (q^2,p^2, p'^2) \nonumber\\
&&=  \Delta^{-1}(p') - \Delta^{-1}(p).
\eea
In particular, for the case of real photons (i.e. $q^2=0$) and for the half-off-shell form factor, namely,  
 the final state being on-mass shell $p'^2=m^2_\pi$ with  $\Delta^{-1}(p')=0$,
 one finds from Eq.~(\ref{WTI1}) that
\bea\label{WTI2}
\Delta^{-1}(p) &=& (p^2 - m^2_\pi) G_1 (0,p^2 , m^2_\pi) 
\nonumber\\
&=&  (p^2 - m^2_\pi) G_1 (0,m^2_\pi, p^2 ).
\eea
Thus, the form factor normalization $G_1(0, m^2_{\pi}, m^2_{\pi})=1$, which can be interpreted as the charge
of the pion,
is attained in the on-shell limit $(p^2=m^2_\pi$) of the initial state
since ${\rm lim}_{p^2 \to m^2_\pi}[(p^2 - m^2_\pi)\Delta(p)]^{-1}=1$.  
However,  the extension to
$G_1(0, m^2_{\pi}, p^2)=1$ for the half-off-shell case ($p^2 \neq m^2_\pi$)
is in general not possible 
due to the nonvanishing  $\Pi(p^2)$ term.
It is also interesting to note that $G_1 (q^2,p^2, p'^2) =G_1 (q^2,p'^2, p^2)$ and $G_2 (q^2,p^2, p'^2) =- G_2 (q^2,p'^2, p^2)$, respectively,
from Eq.~(\ref{WTI1}) and the time-reversal invariance of the strong interaction.

From Eq.~(\ref{WTI1}),
the off-shell form factor $G_1(q^2, p^2, p'^2)$
in the real photon limit~($q^2=0$) is given by
\be\label{WTI3}
 G_1 (0,p^2, p'^2) = \frac{ \Delta^{-1}(p') - \Delta^{-1}(p)}{ p'^2 - p^2}.
\ee
Substituting Eq.~(\ref{WTI3}) back into Eq.~(\ref{WTI1}),  one obtains
\be\label{WTI4}
G_2 (q^2,p^2, p'^2) =\frac{(p'^2 - p^2)[ G_1(0, p^2, p'^2) - G_1(q^2, p^2, p'^2)]}{q^2}.
\ee
In the case of the pion initial state being off-mass-shell but the final state being on-mass-shell,~i.e. $p^2=t$ and $p'^2 =m^2_\pi$,
Eq.~(\ref{WTI4}) becomes~\cite{Rudy}
\be\label{eq4}
 F_2(Q^2,t)=\frac{t-m^2_\pi}{Q^2}\,\left[ F_1(0,t) -F_1(Q^2,t)\right] \, ,
\ee
where $F_i(Q^2,t)\equiv G_i(q^2, t, m^2_\pi) (i=1,2)$ and $Q^2=-q^2$.
We note that $F_2 (Q^2, t)=0$ if both initial and final pions are on-mass-shell (i.e. $p^2=p'^2=m^2_\pi$),
which is consistent with the antisymmetric property of $G_2$, i.e. 
$G_2 (Q^2, p^2, p'^2) = -G_2(Q^2, p'^2, p^2)$.
The normalization of $F_1$ is fixed by requiring $F_1(Q^2=0,t=m^2_\pi)=1$ as we discussed earlier.
The renormalized pion self-energy $\Pi(t)$ is also related to the off-shell pion form factor $F_1(Q^2=0,t)$
as $\Pi(t) = (t-m^2_\pi)\left[1-F_1(0,t)\right]$, assuring the on-mass-shell condition $\Pi(t=m^2_\pi) = 0$
mentioned earlier. We have checked the chiral perturbation theory up to one loop~\cite{Rudy} and confirmed 
that the off-shell pion form factors obtained in Ref.~\cite{Rudy} satisfy
the general formula given by  Eq.~ (\ref{eq4}), as it should be.

From Eqs.~(\ref{eq1}) and~(\ref{eq4}), the half-on-shell ($p'^2=m^2_\pi$) and half-off-shell ($p^2=t<0$) pion-photon vertex   
can be effectively given by
\be \label{eq6}
 \Gamma_\mu = (p'+p)_\mu~F_1 (Q^2,t)
 + q_\mu \frac{(t-m^2_\pi)}{Q^2}~[F_1(0,t) -F_1(Q^2,t)].
\ee
In the elastic electron scattering, the contraction of the second term in Eq.(\ref{eq6}) with the electron current
vanishes due to the current conservation. 
It suggests that $F_2(Q^2,t)$ given by Eq.(\ref{eq4}) cannot be directly measured in the electroproduction process 
due to the transversality of the electron current. 
We note, however, that the ratio of $F_2(Q^2,t)$ to $t-m^2_\pi$ is
nonzero in the limit of $t\to m^2_\pi$ although $F_2(Q^2,t)$ itself goes to zero as $t\to m^2_\pi$. 
To exhibit this more clearly, let us define the new form factor
\begin{equation}\label{eq8}
g(Q^2,t)\equiv \frac{F_2(Q^2,t)}{t-m^2_\pi}\, .
\end{equation}
Then, the off-shell form factor sum rule given by Eq.~(\ref{eq4})  can be rewritten as
\begin{equation}\label{eq7}
F_1(Q^2,t)- F_1(0,t)+ Q^2 g(Q^2,t)=0\, .
\end{equation}
Taking the derivative of Eq.~(\ref{eq7}) with respect to  $Q^2$, one finds the following evolution equation:
\begin{equation}\label{eq9}
\frac{\partial}{\partial Q^2}F_1(Q^2,t)\,+\,g(Q^2,t)+ Q^2\frac{\partial g(Q^2,t)}{\partial Q^2}\,=\,0 \,.
\end{equation}
We should note that  $g(Q^2=0, t=m^2_\pi)$ is associated with the charge radius of the pion elastic form factor.  In other words, since
\begin{equation}\label{eq10}
g(Q^2=0,m^2_\pi) = -\frac{\partial}{\partial Q^2}F_1(Q^2=0,m^2_\pi) =  \frac16 \langle r^2_\pi\rangle
\end{equation}
in the on-mass-shell  limit $t=m^2_\pi$ and at $Q^2=0$, we get the on-mass-shell solution for $g(Q^2,t)$
\begin{equation}\label{eq11}
g(Q^2,m^2_\pi) = \frac16 \langle r^2_\pi \rangle\, +\,  \alpha\; Q^2 +\, \cdots \, ,
\end{equation}
where $\alpha$ is determined by expanding $\frac{\partial}{\partial Q^2}F_1(Q^2,t)$ and $\frac{\partial}{\partial Q^2}g(Q^2,t)$ 
in $Q^2$  
around $Q^2=0$.  
Effectively, the master equation given by Eq.(\ref{eq7}) allows us to extract 
both off-shell form factors simultaneously while the electroproduction process  cannot directly measure $F_2(Q^2,t)$.
Interestingly, however, neither of the two form factors $F_1(Q^2,t)$ and $g(Q^2,t)$ vanishes even in the on-mass-shell limit $t=m^2_\pi$. 

Furthermore, we can continue elaborating  the master equation given by Eq.~(\ref{eq7}), taking the derivative in $t$,
\begin{equation}\label{eq12}
\frac{\partial}{\partial t}F_1(Q^2,t)\,-\, \frac{\partial F_1(0,t)}{\partial t}\,+ Q^2\frac{\partial g(Q^2,t)}{\partial t}\,=\,0 \,,
\end{equation}
and the master equation given by Eq.~(\ref{eq9}), taking the derivative in $t$,
\begin{equation}\label{eq13}
\frac{\partial^2}{\partial t\partial Q^2}F_1(Q^2,t)\,+\, \frac{\partial g(Q^2,t)}{\partial t}\,+ Q^2\frac{\partial^2 g(Q^2,t)}{\partial t\partial Q^2}\,=\,0 \,.
\end{equation}
The form factor $g(Q^2, m^2_\pi)$ is the new observable in the on-mass-shell limit besides the usual charge form factor $F_1(Q^2, m^2_\pi)$
and should be measurable in the experiment of pion electroproduction. In the next section, we shall explicitly show all those properties 
of the off-shell pion form factors using the exactly solvable manifestly covariant model.

\section{Manifestly Covariant Model Calculation}
\label{pionhosffmodel}
\subsection{Model description: Theory}

\begin{figure}
\begin{center}
\includegraphics[height=3cm, width=4cm]{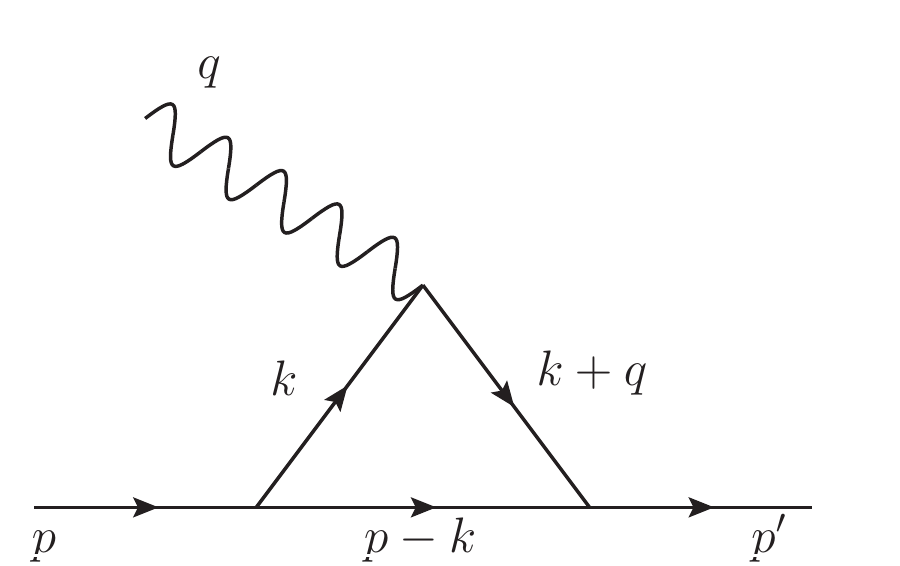}
\caption{\label{fig3}Feynman triangle diagram for the pion off-shell form factors.}
\end{center}
\end{figure}

The vertex function for the initial off-shell ($p^2=t$) and final on-shell ($p'^2=m^2_\pi$) 
$q{\bar q}$ bound-state pion coupled to the virtual photon with the 4-momentum $q$
in the fermion field theory  can be calculated using the tree-level diagram (see Fig.~\ref{fig3})
as
\be\label{Con1}
\Gamma^\mu = i N_c g^2_{\pi q{\bar q}} \int\frac{d^4k}{(2\pi)^4}
\frac{S^\mu}
{N_k N_{k+q} N_{p-k}},
\ee
 where $N_c$ is the number of colors and 
  $g_{\pi q{\bar q}}$ corresponds to the coupling constant of the $\pi q{\bar q}$ vertex.
The denominators
 $N_k = k^2 - m^2_q + i\epsilon$, $N_{k+q}=(k+q)^2 - m^2_q + i\epsilon$, and
$N_{p-k}=(p-k)^2 - m^2_q + i\epsilon$ come from the intermediate quark and antiquark propagators 
with the constituent quark mass
$m_q=m_{\bar q}$, respectively.
The trace term $S^\mu$ in Eq.~(\ref{Con1}) is given by
\be\label{Con2}
 S^\mu
 = {\rm Tr}[\gamma_5(\not\!k +\not\!q + m_q)\gamma^\mu(\not\!k+m_q)\gamma_5(\not\!k-\not\!p+m_q)],
\ee
and the explicit calculation of the triangle loop using the Feynman parametrization and the dimensional
regularization in $d(=4-2\epsilon)$-dimensions is summarized in Appendix.  

From the definition of $\Gamma^\mu = (p'+ p)^\mu F_1(Q^2, t) + q^\mu F_2 (Q^2, t)$,
we then obtain the two form factors $F_1(Q^2,t)$ and $F_2(Q^2,t)$ 
as
\bea\label{off3}
F_1 (Q^2, t) &=& -\frac{N_c g^2_{\pi q{\bar q}}}{8\pi^2} 
\int^1_0\;dx \int^x_0\;dy  
\nonumber\\
&&\times
\biggl[ (1 + 3 y) \biggl( \gamma -\frac{1}{\epsilon} + \frac{1}{2} + {\rm Log}C \biggr) + \frac{\alpha}{C}\biggr],
\nonumber\\
\eea
and
\bea\label{off3-2}
F_2(Q^2, t) &=& -\frac{N_c g^2_{\pi q{\bar q}}}{8\pi^2}
\int^1_0\;dx \int^x_0\;dy 
\nonumber\\
&&\times
\biggl[ 3(1 - 2x + y)  {\rm Log}C + \frac{2\beta-\alpha}{C}\biggr],
\eea
where $\gamma\simeq 0.577$ is the Euler-Mascheroni constant and
\bea\label{Con8}
\alpha &=& (1 +y) (E^2 - m^2_q) - q\cdot E 
+ 2 y p\cdot E -y q\cdot p,
\nonumber\\
\beta &=& (1-x+y) (E^2 - m^2_q) +  (1-2x + 2y) p\cdot E  
+ (x-y)  q\cdot p,
\nonumber\\
\eea
and $C$ and $E$ are given in Appendix \ref{loop-calculation}.
We should note that the form factor $F_2(Q^2,t)$ is free from the UV divergence since the integration
of $(1-2x +y)$ multiplied by the constant factor $(\gamma-1/\epsilon + 1/2)$  gives zero in Eq.~(\ref{off3-2}).

On the other hand,  the form factor $F_1(Q^2,t)$ at $(Q^2=0, t=m^2_\pi)$ is obtained as
\bea\label{F10mpi}
F_1(0,m^2_\pi) &=&  -\frac{N_c g^2_{\pi q{\bar q}}}{8\pi^2}\biggl[
 {\rm Log} (m^2_q)  + \gamma- \frac{1}{\epsilon} - \frac{7}{6}
 \nonumber\\
&&- \frac{2(m^2_\pi - 2m^2_q)}{m_\pi\sqrt{4 m^2_q-m^2_\pi}}\tan^{-1}\biggl( \frac{m_\pi}{\sqrt{4 m^2_q - m^2_\pi} }\biggr)
\biggr].
\nonumber\\
\eea
As the loop correction to the charge form factor $F_1(Q^2,t=m^2_\pi)$ 
must vanish at $Q^2=0$, the charge at $Q^2=0$ is given by a subtraction to the contribution by the loop integral. 
We thus redefine the renormalized charge form factor as 
\be
F^{\rm ren}_1(Q^2,t) = 1+ [ F_1(Q^2,t) - F_1(0, m^2_\pi) ],
\ee
where the loop correction in the square bracket vanishes at $Q^2=0$ and $t=m_\pi^2$
and the normalization of the electric charge is fixed by $F^{\rm ren}_1(0, t=m^2_\pi)=1$.

In this subtractive charge renormalization, the coupling constant $g_{\pi q{\bar q}}$ is still arbitrary and could become a free parameter
to find the best fit for the form factors  from the point of view of the phenomenological application. 
The coupling constant $g_{\pi q{\bar q}}$ is, however, related to the pseudoscalar coupling of the pion {\it vis-\'a-vis} Partially Conserved Axial Current. 
Indeed, the coupling $g_{\pi q{\bar q}}$ may be determined from the comparison of the on-shell pion decay constant 
$f_\pi$ defined by
\be\label{eq:b1}
\la 0|{\bar q}\gamma^\mu\gamma_5 q|\pi(p)\ra
= if_{\pi} p^\mu,
\ee
where $f_\pi$ is obtained by the same model as 
\bea\label{eq:b2}
f_\pi &=& - \frac{N_c g_{\pi q{\bar q}}}{4\pi^2} m_q 
\biggl[ \gamma- \frac{1}{\epsilon} - \frac{3}{2}  + {\rm Log}(m^2_q) 
\nonumber\\
&&+ \frac{2}{m_q} \sqrt{4m^2_q - m^2_\pi} \tan^{-1}\biggl( \frac{m_\pi}{\sqrt{4 m^2_q - m^2_\pi} }\biggr)
\biggr].
\eea
Dividing Eq.~(\ref{F10mpi}) by Eq.~(\ref{eq:b2}),  we obtain 
\be\label{eq:b3}
\frac{g_{\pi q{\bar q}}}{2m_q} = \frac{F_1(0, m^2_\pi)}{f_\pi} + {\cal O}(\epsilon).
\ee
This may motivate to relate $g_{\pi q{\bar q}}$ with $m_q$ and $f^{\rm Exp}_\pi=130$ MeV 
by taking the right hand side of Eq.~(\ref{eq:b3}) as $\frac{F_1^{\rm ren}(0,m^2_\pi)}{f_\pi^{\rm Exp}}$ with $F_1^{\rm ren}(0,m^2_\pi)=1$.
In our numerical calculation, however, 
we take $g_{\pi q{\bar q}}$ as another free parameter in addition to $m_q$ for the best fit of the model calculation compared to the 
experimental data and examine whether the attained value of $g_{\pi q{\bar q}}$ is consistent with the value of $2 m_q/f^{\rm Exp}_\pi$. 

\subsection{Model description: Numerical results}
 \label{resuls-1}

\begin{figure}
\begin{center}
\includegraphics[height=7cm, width=7cm]{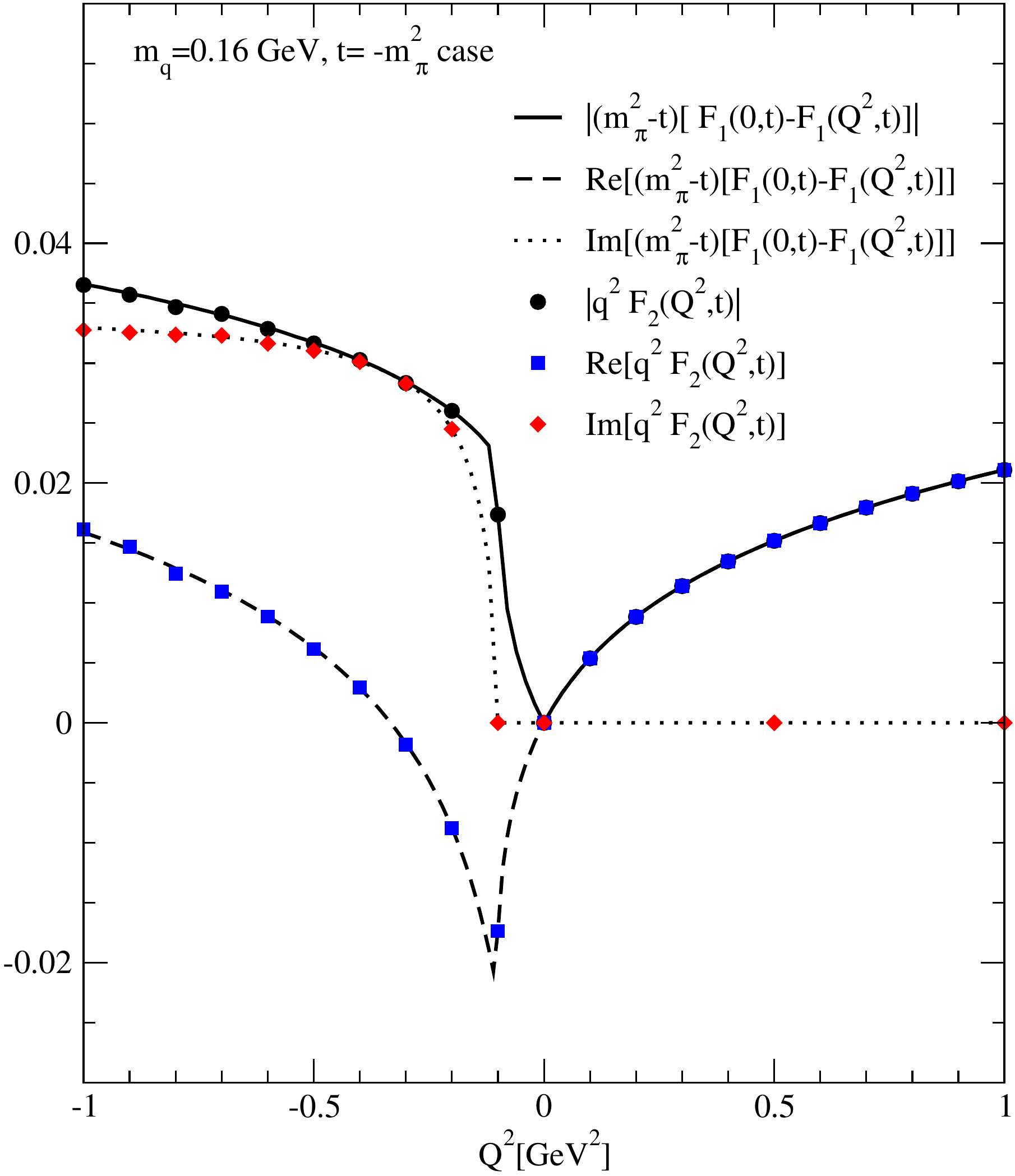}
\caption{\label{fig4} Proof of the WTI given by Eq.~(\ref{eq4}) for the off-shell $\pi^+$ 
obtained from $m_q=0.16$ GeV and $g_{\pi q{\bar q}}=1.11 (2 m_q /f^{\rm Exp}_\pi)$
with fixed $t=-m^2_\pi$ value for $-1 \leq Q^2 \leq 1$ GeV$^2$. }
\end{center}
\end{figure}

The exactly solvable model with the half-off-shell form factors given by  Eqs.~ (\ref{off3}) and (\ref{off3-2}) 
is quantitatively explored in this subsection.
In our numerical calculation, we tried to find the best fits of the form factor $F^{\rm ren}_1(Q^2,t)$ compared
to the experimental data $F^{\rm Exp}_1(Q^2,t)$ for both the off-shell pion ($t\neq m^2_\pi$) (see Table~\ref{t1}) and the on-shell pion ($t=m^2_\pi$) 
as we shall show in Fig.~\ref{fig9}
by adjusting our model parameters $(m_q, g_{\pi q{\bar q}})$.
We found the optimum ranges of quark masses, $0.12\leq m_q \leq 0.16$ GeV,
and the best-fit for the coupling constants, 
$g_{\pi q{\bar q}}=(1.32, 1.20, 1.11)  (2 m_q /f^{\rm Exp}_\pi)$
for $m_q =(0.12, 0.14, 0.16)$ GeV, respectively. That is, our phenomenological best fit coupling constants $g_{\pi q{\bar q}}$ are not
much different from the values of $2 m_q /f^{\rm Exp}_\pi$ and we check the sensitivity of the coupling $g_{\pi q{\bar q}}$
for given quark mass as we shall show in Fig.~\ref{fig9}.
From now on, we shall denote our results for the renormalized form factor $F^{\rm ren}_1(Q^2,t)$ as $F_1(Q^2,t)$ for convenience.

In Fig.~\ref{fig4}, we provide the explicit proof of the WTI given by Eq.~(\ref{eq4}) with the two off-shell form factors 
$F_1$ and $F_2$ computed independently using 
$m_q=0.16$ GeV and $g_{\pi q{\bar q}}=1.11 (2 m_q /f^{\rm Exp}_\pi)$
with a fixed $t=-m^2_\pi$ value
for $-1 \leq Q^2 \leq 1$ GeV$^2$. Note here that 
we cover both timelike $(Q^2=-q^2<0)$ and spacelike $(Q^2>0$) regions.
The timelike result is obtained from the analytic continuation by changing $Q^2$ to $-Q^2$ in the form factors 
of the spacelike region and vice versa. The solid, dashed, and dotted lines 
represent  the results of $|(m_{\pi}^2 - t) [F_1(0,t) -  F_1(Q^2, t)] |$, ${\rm Re}[(m_{\pi}^2 - t) [F_1(0,t) -  F_1(Q^2, t)] ]$,  and
${\rm Im}[(m_{\pi}^2 - t) [F_1(0,t) -  F_1(Q^2, t)] ]$, respectively. 
Our independent calculations of 
$|q^2 F_2 (Q^2,t)|$ (circle), ${\rm Re}[q^2 F_2 (Q^2,t)]$ (square) and ${\rm Im}[q^2 F_2 (Q^2,t)]$ (diamond) shown
in Fig.~\ref{fig4}
prove explicitly  that our model calculation satisfies the WTI given by Eq.~(\ref{eq4}).

The kink in Fig.~\ref{fig4}  of the timelike region is the point where the threshold starts at 
$q^2 = 4 m^2_q$. At $q^2 = 4 m^2_q$, the imaginary parts of the form factors start to develop, where  the $q\bar q$ continuum begins in the model.
Although our analytic covariant model is too simple to illustrate the timelike region $Q^2 < 0$ lacking the 
more realistic feature of the vector meson resonances observed experimentally (see e.g. Refs.~\cite{PedlarPRL05,SethPRL13}),
it may provide at least a theoretical tool to discuss the off-mass-shell aspect of the charged pion form factors 
involved in the electroproduction process, satisfying the master equation given by Eq.(\ref{eq7}) derived from the general WTI given by Eq.(\ref{eq2}). 

\begin{figure*}
\begin{center}
\includegraphics[height=4cm, width=5cm]{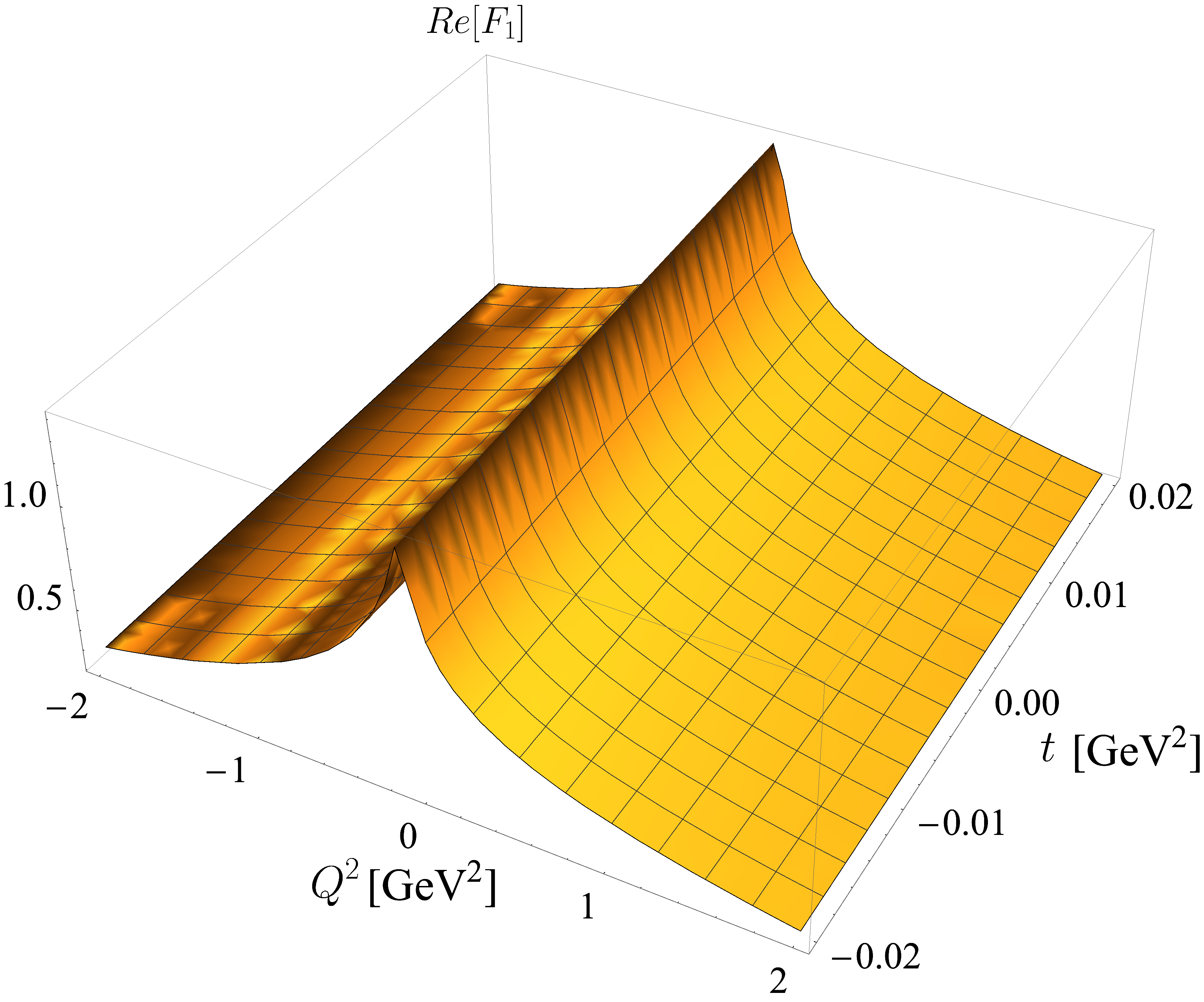}
\includegraphics[height=4cm, width=5cm]{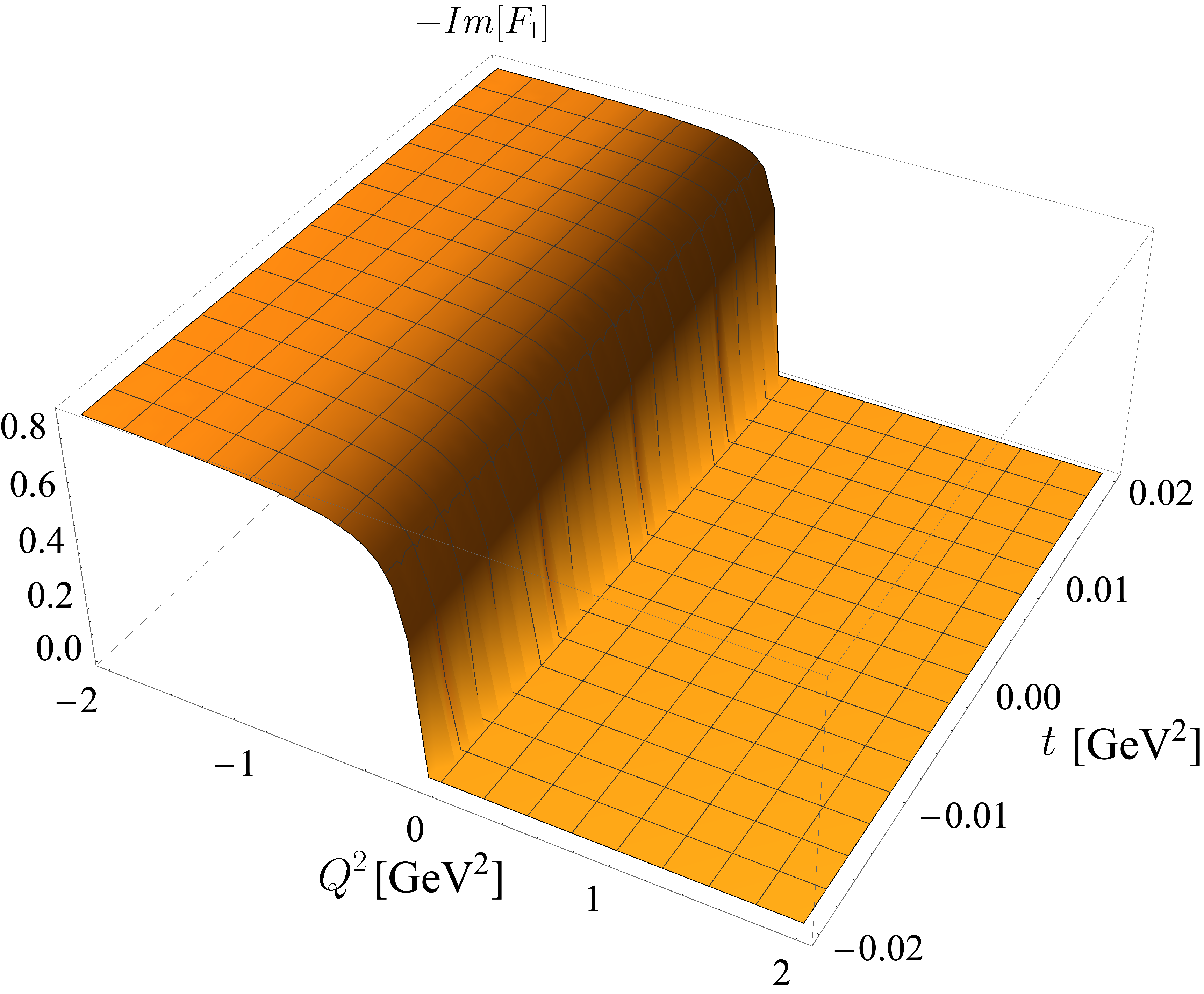}
\includegraphics[height=4cm, width=5cm]{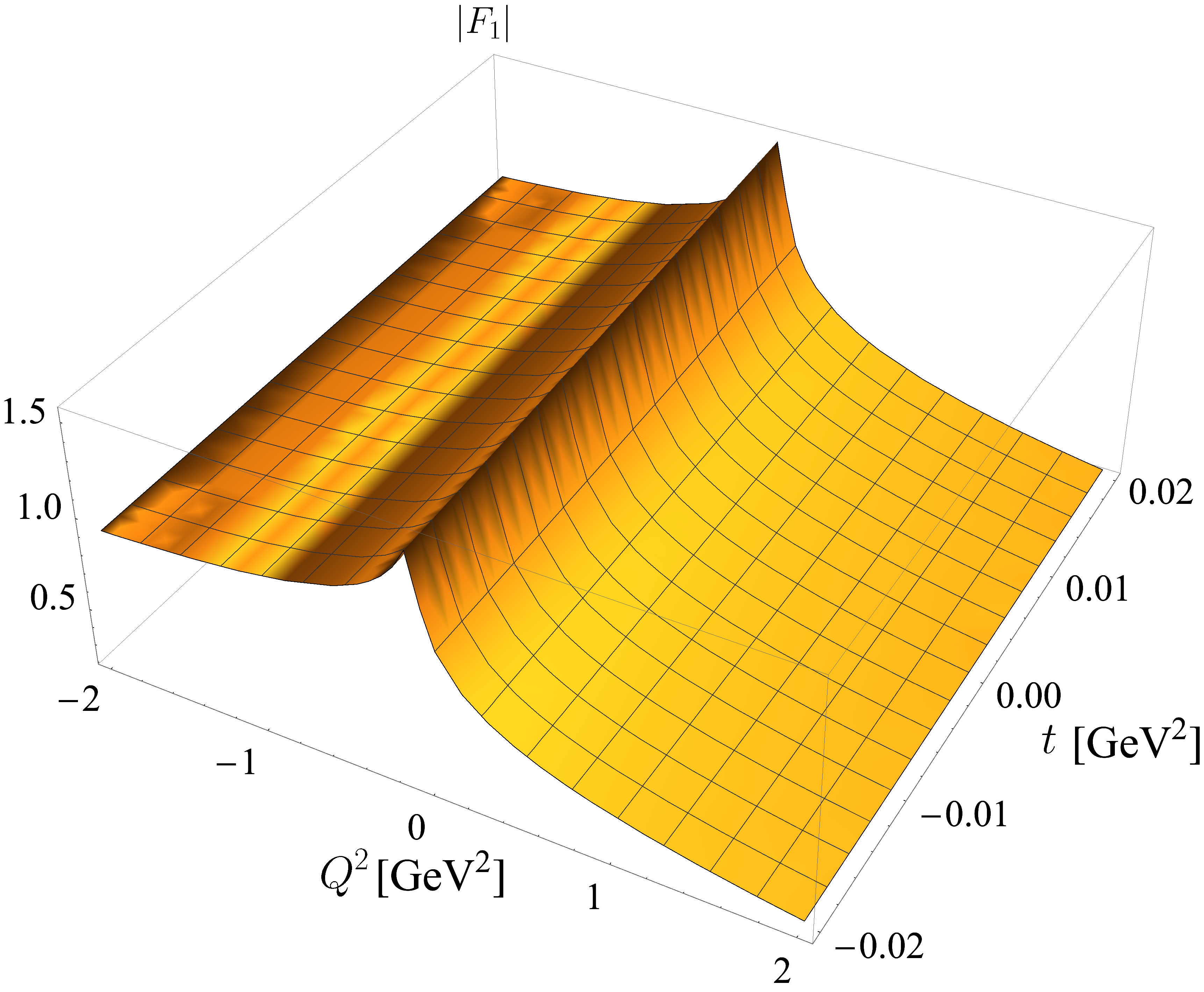}
\vspace{.5cm}
\includegraphics[height=4cm, width=5cm]{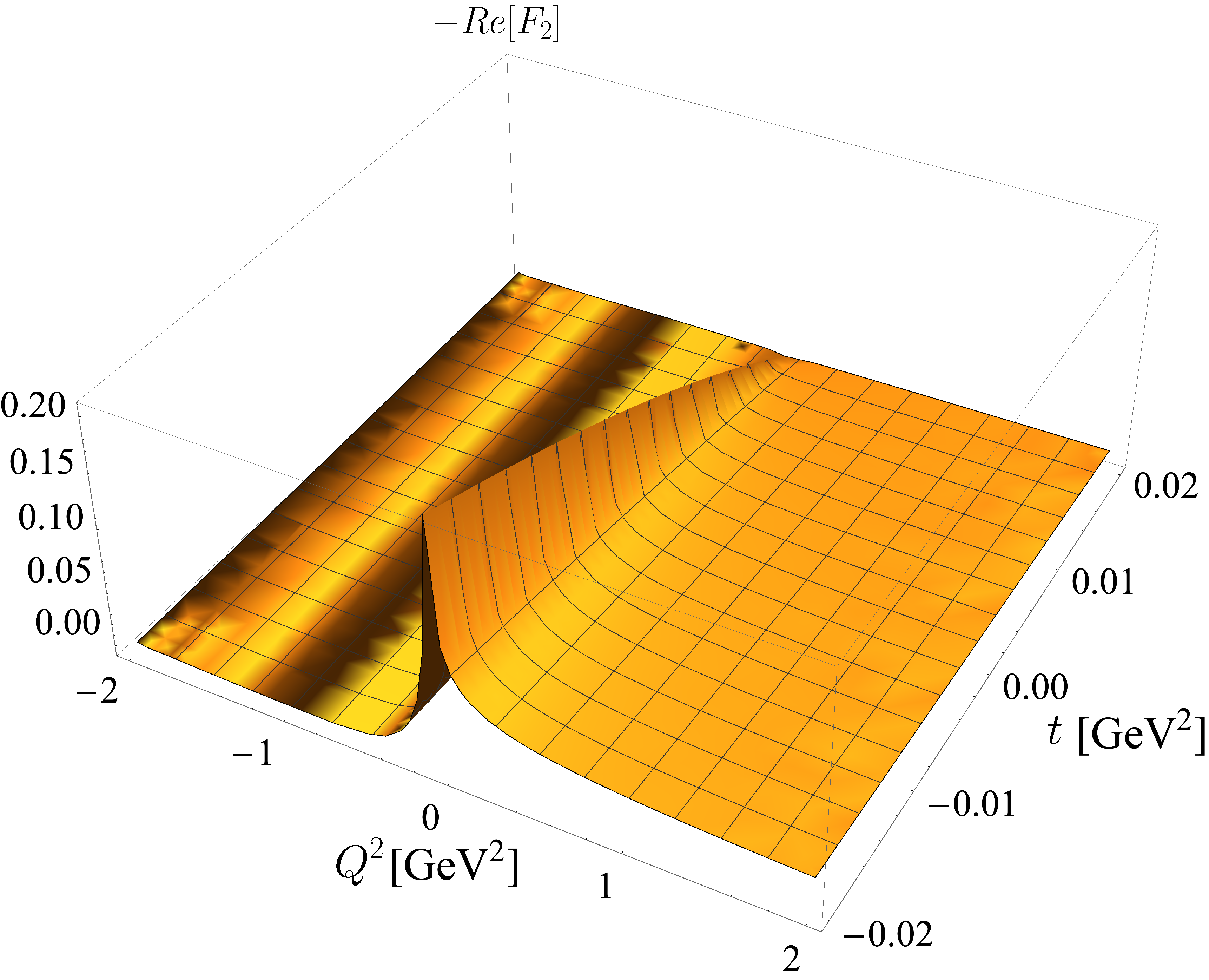}
\includegraphics[height=4cm, width=5cm]{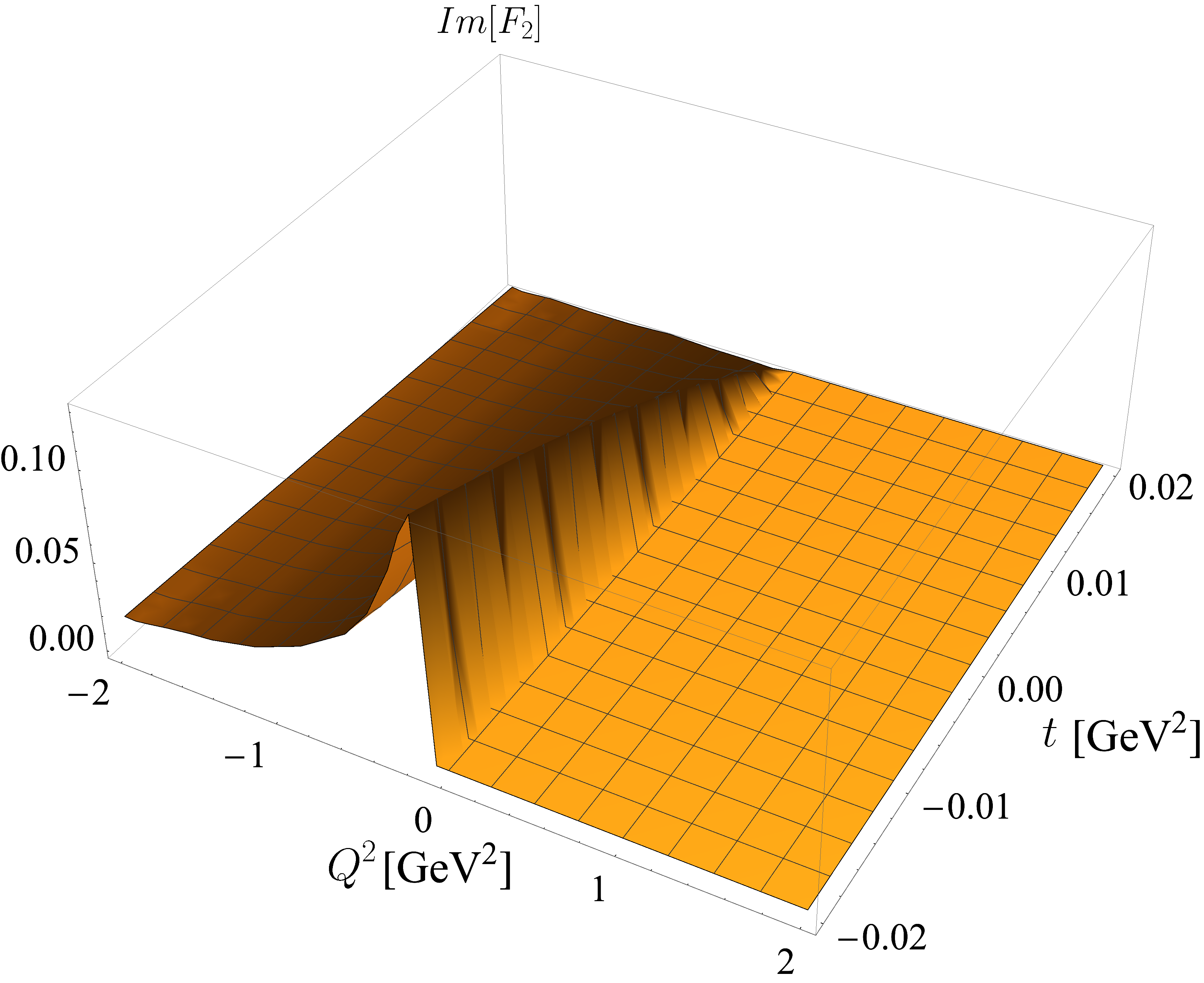}
\includegraphics[height=4cm, width=5cm]{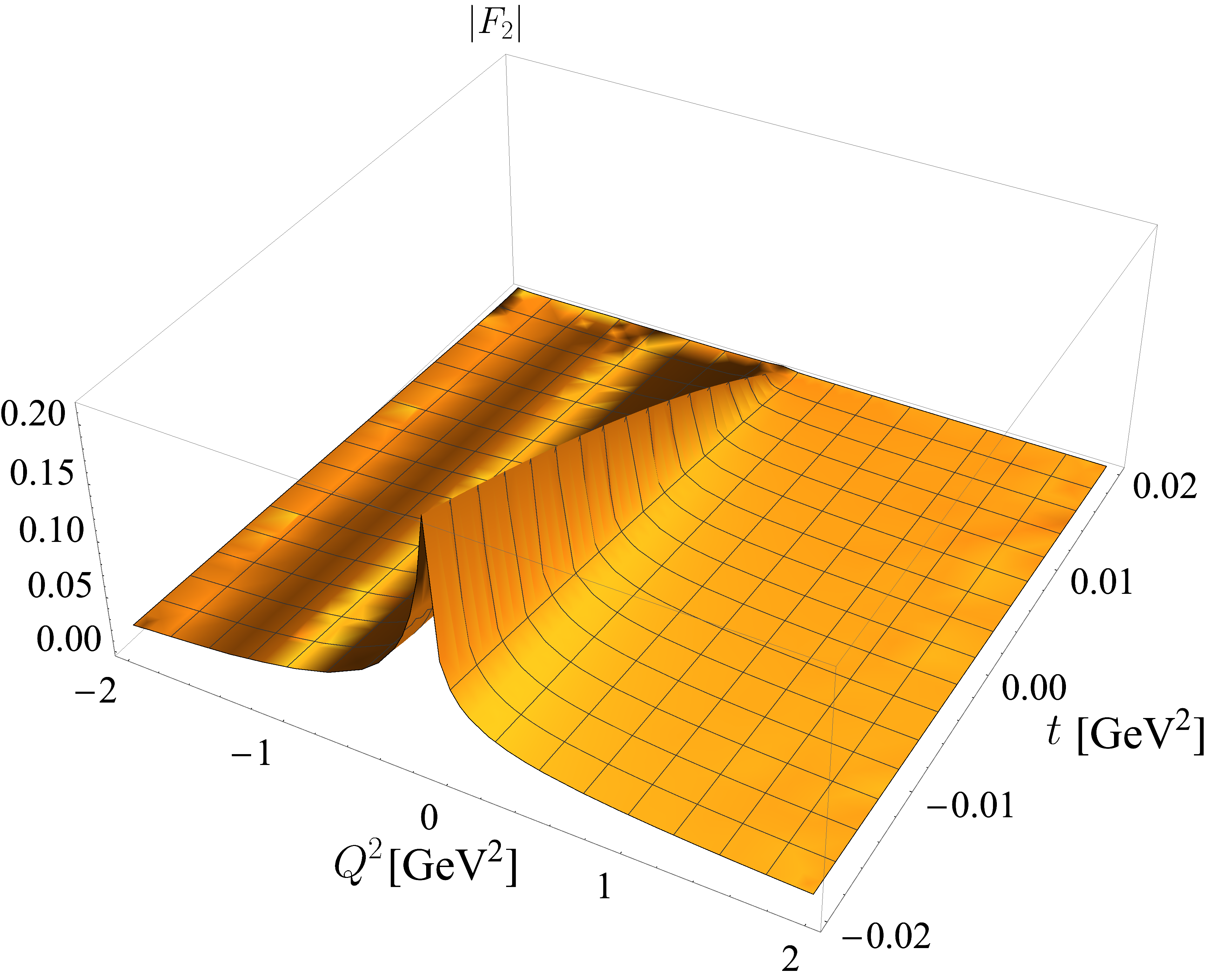}
\caption{\label{fig5} The 3D plots of $F_1(Q^2, t)$ (upper panel) and $F_2(Q^2, t)$ (lower panel) 
for $-2\leq Q^2\leq 2$ GeV$^2$ and $-m^2_\pi \leq t\leq m^2_\pi$ GeV$^2$. Left, middle, and right panels represent
the results of Re[$F_{i}$],  Im[$F_i$], and $|F_i|$ ($i=1,2$), respectively. 
The used model parameters are $m_q=0.16$ GeV and $g_{\pi q{\bar q}}=1.11 (2 m_q /f^{\rm Exp}_\pi)$.
}
\end{center}
\end{figure*}

The overall landscape of the half-on-shell form factors, $F_1(Q^2, t)$ and $F_2(Q^2, t)$,  
obtained from $m_q=0.16$ GeV and $g_{\pi q{\bar q}}=1.11 (2 m_q /f^{\rm Exp}_\pi)$
for both spacelike and timelike regions 
are shown in Fig.~\ref{fig5}, in which the modulus and the real and  imaginary parts are presented. 
The figure shows the 3D plots of $F_1(Q^2, t)$ (upper panel) and $F_2(Q^2, t)$ (lower panel)
for $-2\leq Q^2\leq 2$ GeV$^2$ and $-m^2_\pi \leq t\leq m^2_\pi$ GeV$^2$. Left, middle, and right panels represent
the results of Re[$F_{i}$],  Im[$F_i$], and the modulus $|F_i|=\sqrt{ ({\rm Re}[F_i])^2 + ({\rm Im}[F_i])^2}$ ($i=1,2$), respectively.  
The imaginary parts of both $F_1$ and $F_2$ start to appear at $q^2=4 m^2_q$ regardless of the off-shell value $t$.
For the form factor $F_2(Q^2,t)$, it clearly satisfies $F_2(Q^2,t)=0$ at the on-shell limit $t=m^2_\pi$ in accordance with the WTI given by Eq.~(\ref{eq4}).
However, $F_2$ is no longer zero for $t\neq m^2_\pi$ values and shows quite different cusp behavior 
from $F_1$ in the timelike region as $t$ gets away from the on-shell
$t=m^2_\pi$ value.  This may suggest that the different extrapolation methods from $t<0$ to $t=m^2_\pi$  are required for $F_1$ and $F_2$, with
the proviso that the model lacks the more realistic feature of the vector meson resonances observed experimentally in the timelike region. 
Despite this limitation, our results illustrate that it may be possible to extract the two form factors by probing 
different aspects of the pion structure.

\begin{figure}
\begin{center}
\includegraphics[height=4cm, width=4cm]{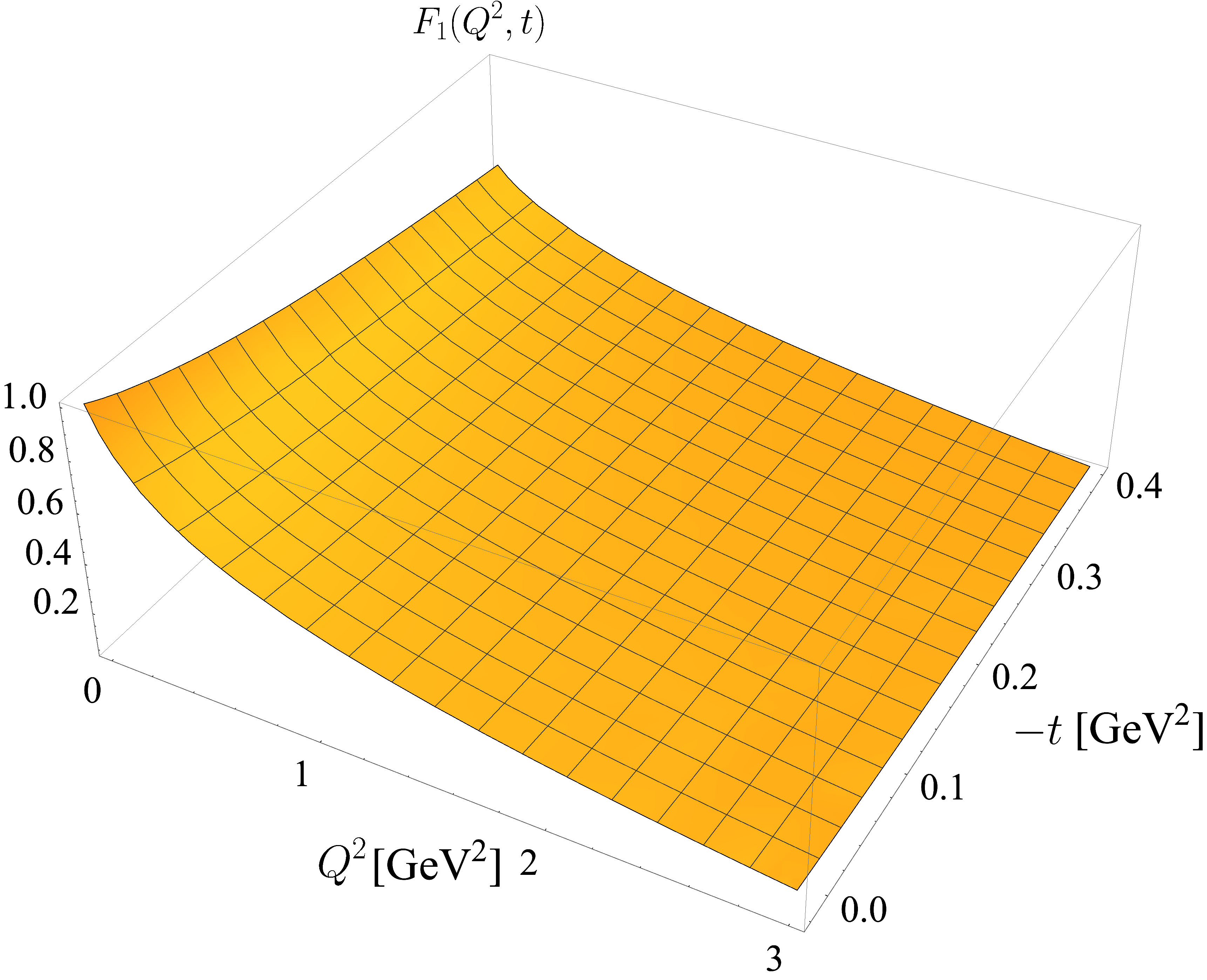}
\includegraphics[height=4cm, width=4cm]{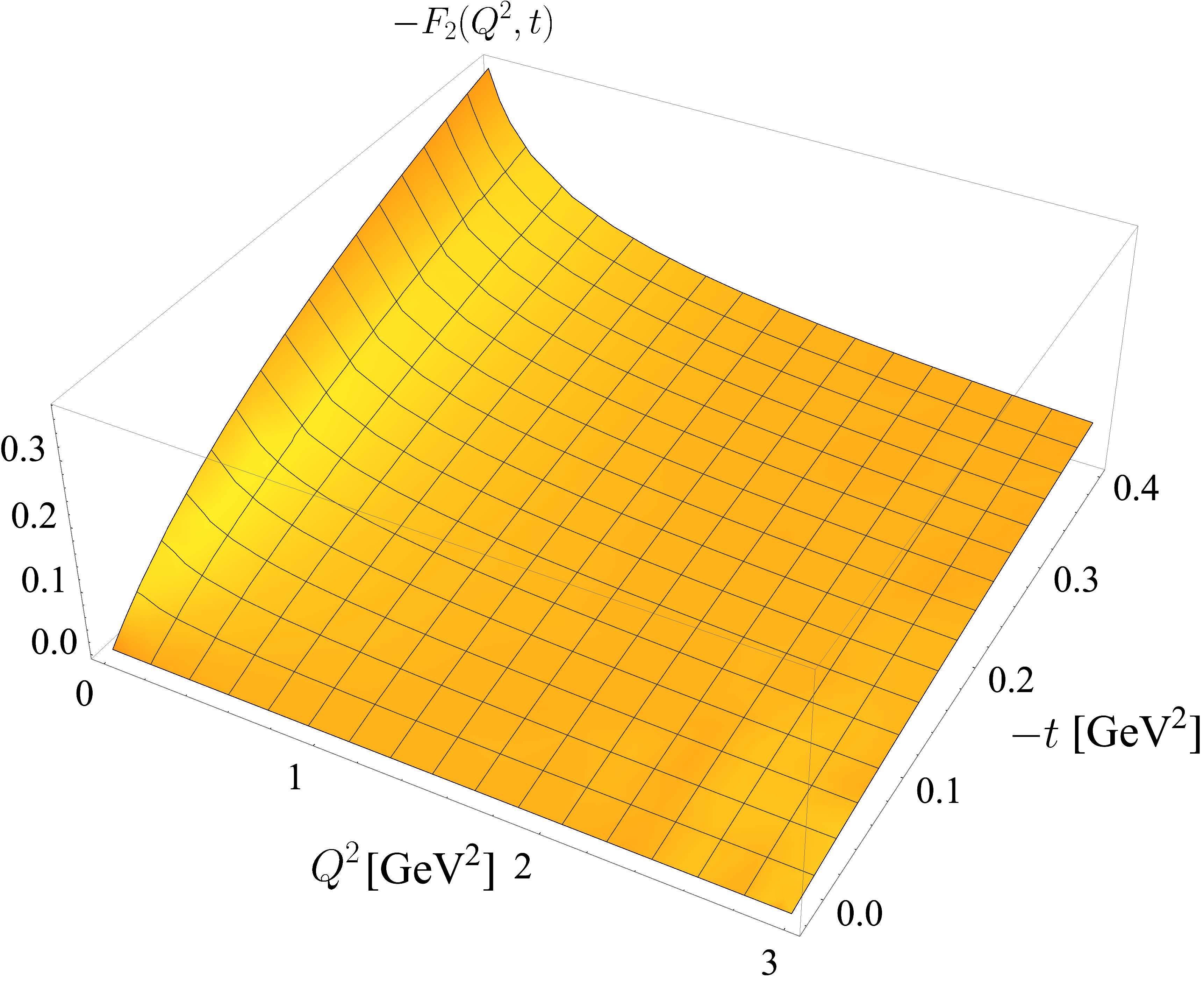}
\includegraphics[height=4cm, width=4cm]{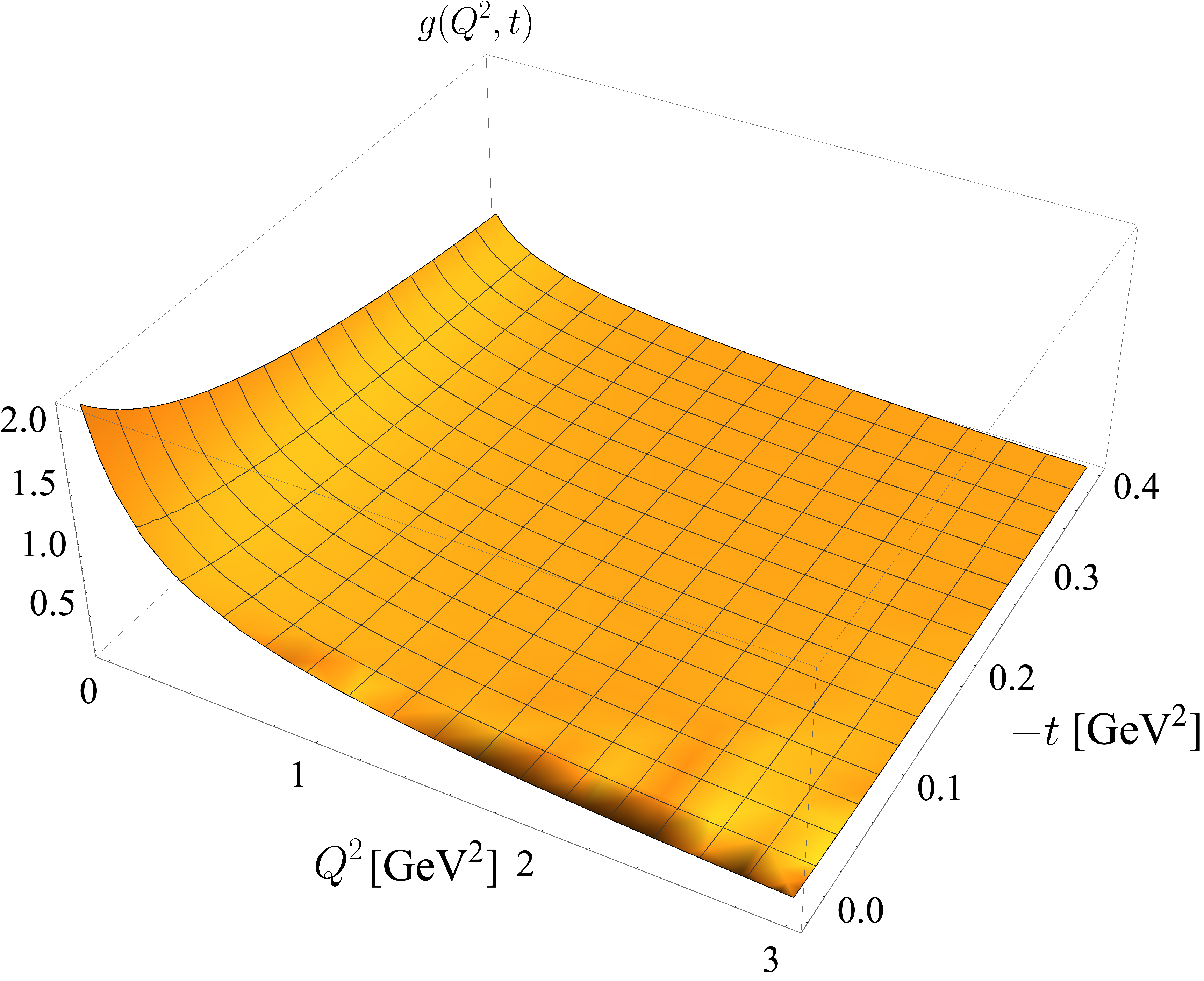}
\includegraphics[height=4cm, width=4cm]{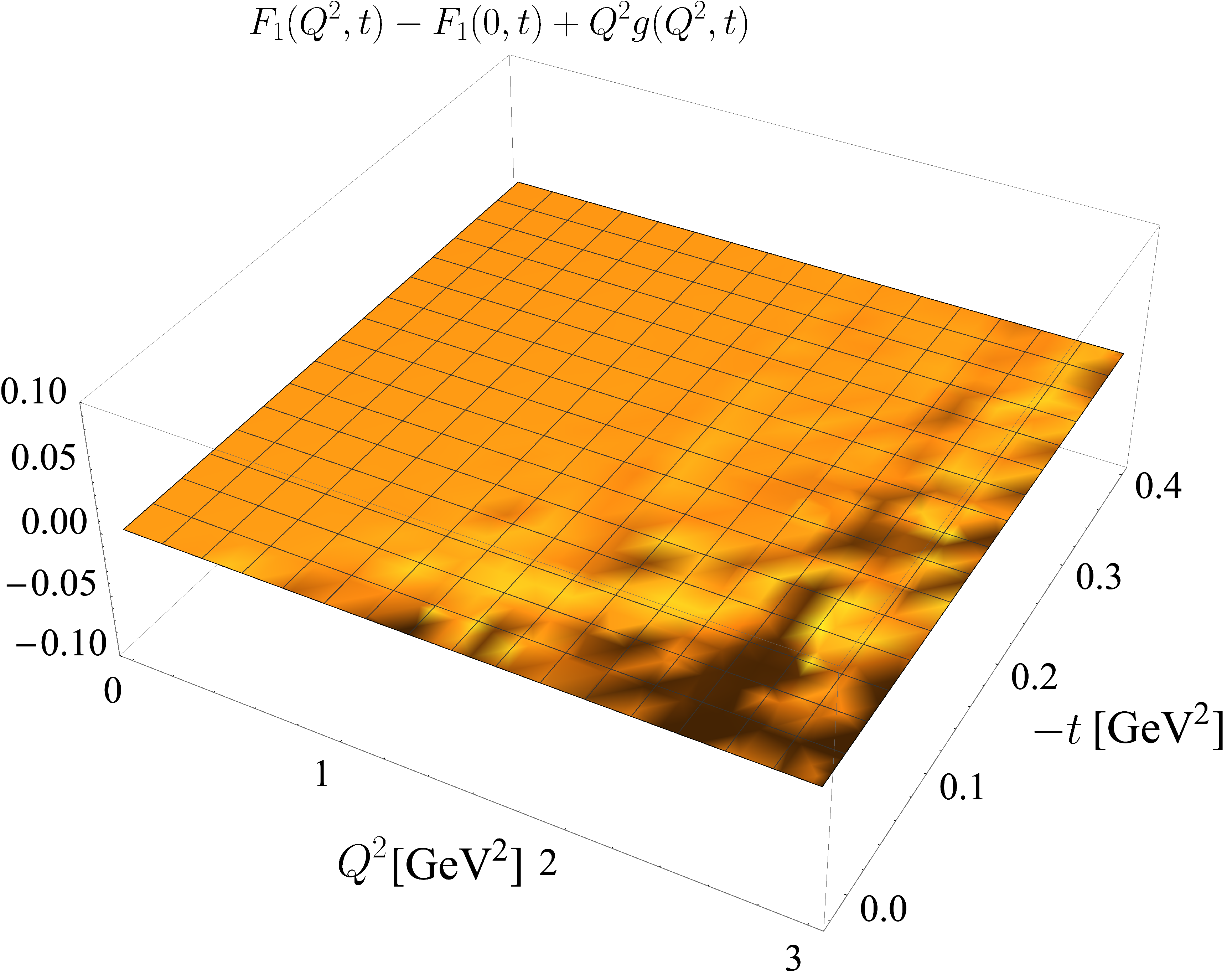}
\caption{\label{fig6} The 3D plots of $F_1(Q^2, t)$ (top left), $F_2(Q^2,t)$ (top right), 
$g(Q^2, t)$ (bottom left) and the sum rule (bottom right) given by Eq.~(\ref{eq7})
for the spacelike momentum transfer region $0\leq Q^2\leq 3$ GeV$^2$ and $m^2_\pi \geq t\geq -0.4$ GeV$^2$.
The used model parameters are $m_q=0.16$ GeV and $g_{\pi q{\bar q}}=1.11 (2 m_q /f^{\rm Exp}_\pi)$.
}
\end{center}
\end{figure}

The landscapes of the half-off-shell spacelike  form factors 
given in Fig.~\ref{fig5}
are shown in more detail in Fig.~\ref{fig6}, as it is relevant
 for our forthcoming analysis of the experimental data. The figure represents the 3D plots of $F_1(Q^2, t)$ (top left), $-F_2(Q^2,t)$ (top right), 
$g(Q^2, t)$ (bottom left) and the master equation (bottom right) given by Eq.~(\ref{eq7})
 for  the momentum transfer region $0\leq Q^2\leq 3$ GeV$^2$ and $m^2_\pi \geq t \geq -0.4$ GeV$^2$. 
While the form factor $F_2(Q^2,t)$ goes to zero as $t\to m^2_\pi$, the form factor $g(Q^2, t)$ is nonzero even in the
on-mass-shell limit. Furthermore,  $F_1(0, t)$ shows some dependencies on $t$, which is necessary to know
in the case of extracting $F_2(Q^2,t)$ from the pion electroproduction data.
 In particular, the value of $g(Q^2=0,t=m^2_\pi)$ corresponds to the charge radius of a pion.
The covariant and analytical model is checked against the fulfillment  of the master equations, i.e. 
the sum rules given by Eqs.~(\ref{eq7}), (\ref{eq9}), (\ref{eq12}) and (\ref{eq13}),
and we display the fulfillment of Eq.~(\ref{eq7}) in the figure (bottom right) as an explicit illustration.
The verification of these master equations  gives not only an
indirect check on the fulfillment  of the  WTI by the model, but also our numerical accuracy.

\section{Extraction of the off-shell form factors from the experimental cross section} \label{xsection}
\subsection{Extraction of half-off-shell pion form factors}\label{results-2}

The off-shell form factor $F_1(Q^2,t)$ can be extracted from the  exclusive cross section
 for  $^1{\rm H}(e,e',\pi^+)n$ in the kinematical region of small $t$, such that the $t$-channel process dominates
near  the pion pole at $t=m^2_\pi$~\cite{Blok2008,Huber2008}. 
To minimize background contributions, it is also necessary to separate out the longitudinal
cross section $\sigma_{\rm L}$, via the Rosenbluth separation 
depending on the polarization states of the virtual 
photon in terms of the longitudinal differential cross section ($d\sigma_{\rm L}/dt$), the transverse differential cross section ($d\sigma_{\rm T}/dt$), 
and the two other differential cross sections due to interference ~($d\sigma_{\rm LT}/dt$ and $d\sigma_{\rm TT}/dt$). 

Since the minimum physical value of $-t$ is nonzero and increases with the increasing value of $Q^2$ and
decreasing value of the invariant mass $W$ of the produced pion-nucleon system, 
more reliable extraction of the on-shell pion form factor $F_\pi(Q^2)=F_1(Q^2,t=m^2_\pi)$ should
be performed at smaller $-t$ and higher $W$ (for a fixed $Q^2$) as discussed in Ref.~\cite{Huber2008}.
 In Ref~\cite{Huber2008}, ample discussions were devoted to the reliability issue of the Chew-Low extrapolation 
method and the use of the Regge model presented in Ref.~\cite{VGL} as well as the encouragement on additional models that one may use for the 
task of the form factor extraction.

The basis of  the Chew-Low method is  the Born-term model formula for the pion-pole 
contribution to $\sigma_{\rm L}$, where the pion-pole contribution to $\sigma_{\rm L}$ is given by
\begin{equation}\label{ex28}
 N \frac{d\sigma_L}{dt}=
4 \hbar c (e G_{\pi N N})^2 
 \frac{-tQ^2}{(t-m^2_{\pi})^2} F^2_\pi(Q^2)~.
\end{equation}
Here, $e^2/(4 \pi \hbar c)=1/137$ and the factor $N$ which depends on the flux factor used in the definition
of $d\sigma_{\rm L}/dt$ is given by
\be\label{ex29}
N=32 \pi \left(W^2 -m^2_p\right)
\sqrt{(W^2-m^2_p)^2 + Q^4 + 2 Q^2 (W^2+ m^2_p)}.
\ee
For the form factor~$G_{\pi NN}(t)$, we follow the usual monopole type of parametrization 
\begin{eqnarray}\label{ex30}
 \ G_{\pi NN } (t) & = & G_{\pi NN}(m^2_\pi) 
\left( \frac{\Lambda^2_\pi - m^2_\pi}{\Lambda^2_\pi ~- ~t} \right),
\end{eqnarray}
where $G_{\pi NN}(m^2_\pi)=13.4$ and $\Lambda_\pi=0.80$ GeV have been taken in the extraction of $F_\pi$
from the Jefferson Lab experiment~\cite{Huber2008}. We use the same values of $G_{\pi NN}(m^2_\pi)$ and $\Lambda_\pi$ in
our numerical extraction of the off-shell form factors $F_1(Q^2,t)$ and $F_2(Q^2,t)$ [or $g(Q^2,t)$].

\begin{table*}[htb!]
 \begin{center}
\caption{Pion form factors extracted from experimental cross section for $d\sigma_{\rm L}/dt$ given in Table VII of Ref.~\cite{Blok2008} 
vs  solvable model with $m_q=0.14\pm0.02$ GeV.  
The coupling constants, 
$g_{\pi q{\bar q}}=(1.32, 1.20, 1.11) (2 m_q /f^{\rm Exp}_\pi)$,
are used for $m_q =(0.12, 0.14, 0.16)$ GeV, respectively. 
$(Q^2, t)$ are in units of GeV$^2$, and $g(Q^2,t)$ is in units of GeV$^{-2}$. 
 }\label{t1}
\bigskip
\begin{tabular}{ccccccc} 
\hline
\hline 
 $Q^2$ & $-t$ & ~~$F^{\rm Exp}_1(Q^2,t)$  & ~$F^{\rm Cov}_1(Q^2,t)$ & ~$F^{\rm Cov}_1(0,t)$ & $g^{\rm Exp}(Q^2,t)$ & $g^{\rm Cov}(Q^2,t)$\\
\hline
            &          &$\la Q^2 \ra =0.60$ GeV$^2$, &  $W=1.95$ GeV                    &                                         &                                            &                              \\
0.526  & 0.026  & 0.502 $\pm$ 0.013    & $0.487^{+0.032}_{-0.039} $& $0.891^{+0.019}_{-0.030}$ & $0.740^{+0.060}_{-0.082}$ & $0.768^{-0.024}_{+0.018}$ \\
0.576  & 0.038  & 0.440 $\pm$ 0.010  & $0.462^{+0.032}_{-0.039}$ & $0.869^{+0.022}_{-0.033}$ & $0.745^{+0.055}_{-0.075}$ & $0.708^{-0.016}_{+0.008}$\\ 
0.612  & 0.050  & 0.413  $\pm$ 0.011  & $0.443^{+0.030}_{-0.038}$  & $0.849^{+0.024}_{-0.036}$ & $0.712^{+0.058}_{-0.076}$ & $0.664^{-0.010}_{+0.003}$\\
0.631  & 0.062  & 0.371 $\pm$  0.014    & $0.430^{+0.030}_{-0.036}$  & $0.831^{+0.026}_{-0.038}$ & $0.729^{+0.063}_{-0.082}$ & $0.635^{-0.007}_{-0.002}$\\
0.646  & 0.074  &  0.340 $\pm$  0.022   & $0.419^{+0.030}_{-0.036}$  & $0.814^{+0.027}_{-0.039}$ & $0.734^{+0.076}_{-0.095}$ & $0.611^{-0.004}_{-0.005}$\\
\hline 
            &          &$\la Q^2 \ra =0.75$ GeV$^2$, &  $W=1.95$ GeV                    &                                         &                                            &                              \\
0.660  & 0.037  & 0.397  $\pm$ 0.019  & $0.435^{+0.030}_{-0.036}$ & $0.870^{+0.023}_{-0.032}$ & $0.717^{+0.063}_{-0.078}$ & $0.660^{-0.012}_{+0.005}$ \\
0.707  & 0.051  & 0.360 $\pm$ 0.017    & $0.414^{+0.030}_{-0.035}$  & $0.848^{+0.024}_{-0.036}$  & $0.690^{+0.058}_{-0.075}$ & $0.613^{-0.006}_{-0.001}$\\
0.753  & 0.065  & 0.358  $\pm$  0.015  & $0.394^{+0.029}_{-0.034}$  & $0.827^{+0.026}_{-0.039}$ & $0.623^{+0.054}_{-0.072}$ & $0.574^{-0.003}_{-0.006}$ \\
0.781  & 0.079  & 0.324  $\pm$  0.018   & $0.381^{+0.027}_{-0.033}$ & $0.807^{+0.028}_{-0.040}$  & $0.618^{+0.059}_{-0.074}$ & $0.546^{-0.001}_{-0.009}$\\
0.794  & 0.093  & 0.325  $\pm $ 0.022   & $0.371^{+0.028}_{-0.032}$  & $0.789^{+0.029}_{-0.041}$ & $0.584^{+0.065}_{-0.079}$  & $0.526^{+0.003}_{-0.011}$\\
\hline 
            &          &$\la Q^2 \ra =1.00$ GeV$^2$, &  $W=1.95$ GeV                    &                                         &                                            &                              \\
0.877  & 0.060  & 0.342 $\pm$ 0.014   & $0.366^{+0.027}_{-0.031}$  & $0.834^{+0.026}_{-0.038}$ & $0.561^{+0.046}_{-0.059}$ & $0.533^{-0.001}_{-0.006}$   \\
0.945  & 0.080  & 0.327 $\pm$ 0.012  & $0.343^{+0.025}_{-0.030}$  & $0.806^{+0.028}_{-0.040}$ & $0.507^{+0.042}_{-0.055}$ & $0.490^{+0.003}_{-0.010}$ \\
1.010  & 0.100  &  0.311 $\pm$ 0.012   & $0.322^{+0.024}_{-0.029}$   & $0.781^{+0.030}_{-0.042}$ & $0.465^{+0.042}_{-0.053}$  & $0.454^{+0.006}_{-0.013}$\\
1.050  & 0.120  & 0.282 $\pm$ 0.016   & $0.307^{+0.023}_{-0.027}$   & $0.758^{+0.031}_{-0.043}$ & $0.453^{+0.045}_{-0.056}$  & $0.430^{+0.007}_{-0.015}$\\
1.067  & 0.140  & 0.233 $\pm$ 0.028 & $0.297^{+0.023}_{-0.026}$    & $0.737^{+0.032}_{-0.043}$ & $0.472^{+0.057}_{-0.066}$ & $0.412^{+0.009}_{-0.015}$\\
\hline 
            &          &$\la Q^2 \ra =1.60$ GeV$^2$, &  $W=1.95$ GeV                    &                                         &                                            &                              \\
1.455  & 0.135  & 0.258 $\pm$ 0.010    & $0.237^{+0.018}_{-0.021}$ & $0.742^{+0.032}_{-0.043}$ & $0.332^{+0.029}_{-0.037}$ &$0.347^{+0.010}_{-0.015}$ \\
1.532  & 0.165  & 0.245 $\pm$ 0.010   & $0.219^{+0.016}_{-0.020}$  & $0.714^{+0.032}_{-0.044}$  & $0.306^{+0.028}_{-0.035}$ & $0.323^{+0.011}_{-0.016}$\\
1.610 & 0.195  & 0.222 $\pm$ 0.012    & $0.201^{+0.015}_{-0.018}$  & $0.688^{+0.033}_{-0.044}$ & $0.289^{+0.028}_{-0.034}$ & $0.302^{+0.012}_{-0.016}$\\
1.664  & 0.225  & 0.203 $\pm$ 0.013   & $0.188^{+0.014}_{-0.017}$  & $0.665^{+0.034}_{-0.045}$ & $0.278^{+0.028}_{-0.035}$& $0.286^{+0.012}_{-0.016}$\\
1.702  & 0.255  & 0.227 $\pm$ 0.016   & $0.177^{+0.014}_{-0.015}$  & $0.644^{+0.034}_{-0.044}$ & $0.245^{+0.029}_{-0.035}$ & $0.274^{+0.012}_{-0.017}$\\
\hline 
            &          &$\la Q^2 \ra =1.60$ GeV$^2$, &  $W=2.22$ GeV                    &                                         &                                            &                           \\
1.416  & 0.079  & 0.270 $\pm$ 0.010   & $0.259^{+0.019}_{-0.022}$ &$0.807^{+0.028}_{-0.040}$  & $0.379^{+0.027}_{-0.035}$ & $0.387^{+0.006}_{-0.012}$\\
1.513  & 0.112  & 0.258 $\pm$ 0.010  & $0.235^{+0.018}_{-0.021}$  & $0.767^{+0.030}_{-0.043}$ & $0.336^{+0.027}_{-0.035}$ & $0.351^{+0.009}_{-0.014}$\\
1.593 & 0.139  & 0.251 $\pm$ 0.010   & $0.217^{+0.016}_{-0.019}$  & $0.738^{+0.032}_{-0.043}$ & $0.306^{+0.026}_{-0.034}$ & $0.327^{+0.010}_{-0.015}$ \\
1.667  & 0.166  & 0.241 $\pm$ 0.012    & $0.201^{+0.015}_{-0.018}$  & $0.713^{+0.033}_{-0.044}$ & $0.283^{+0.027}_{-0.033}$ & $0.307^{+0.011}_{-0.016}$\\
1.763  & 0.215  & 0.200 $\pm$ 0.018   & $0.179^{+0.013}_{-0.017}$ & $0.672^{+0.034}_{-0.044}$ & $0.268^{+0.029}_{-0.035}$ & $0.280^{+0.011}_{-0.017}$\\
\hline
            &          &$\la Q^2 \ra =2.45$ GeV$^2$, &  $W=2.22$ GeV                    &                                         &                                            &                               \\
2.215  & 0.145 &  0.188 $\pm$ 0.008 &  $0.146^{+0.010}_{-0.012}$  & $0.732^{+0.033}_{-0.043}$ & $0.246^{+0.018}_{-0.023}$ & $0.265^{+0.010}_{-0.014}$\\
2.279  & 0.202 & 0.178 $\pm$ 0.008   & $0.129^{+0.009}_{-0.011}$ & $0.682^{+0.034}_{-0.044}$ &  $0.221^{+0.019}_{-0.023}$ & $0.243^{+0.011}_{-0.015}$ \\
2.411  & 0.245 &  0.163 $\pm$  0.009  & $0.109^{+0.008}_{-0.009}$ & $0.650^{+0.037}_{-0.044}$ & $0.202^{+0.019}_{-0.022}$ & $0.224^{+0.011}_{-0.014}$ \\
2.539  & 0.288 & 0.156 $\pm$  0.011 &  $0.092^{+0.006}_{-0.007}$  & $0.622^{+0.034}_{-0.043}$ & $0.184^{+0.017}_{-0.022}$ & $0.209^{+0.011}_{-0.014}$ \\
2.703  & 0.365 &  0.150 $\pm$  0.016  &   $0.068^{+0.004}_{-0.005}$ & $0.579^{+0.033}_{-0.043}$ & $0.159^{+0.018}_{-0.022}$ & $0.189^{+0.011}_{-0.014}$ \\
  \hline
  \hline 
\end{tabular}
\end{center}
\end{table*}

\begin{figure}
\begin{center}
\includegraphics[width=5.5cm,angle=-90]{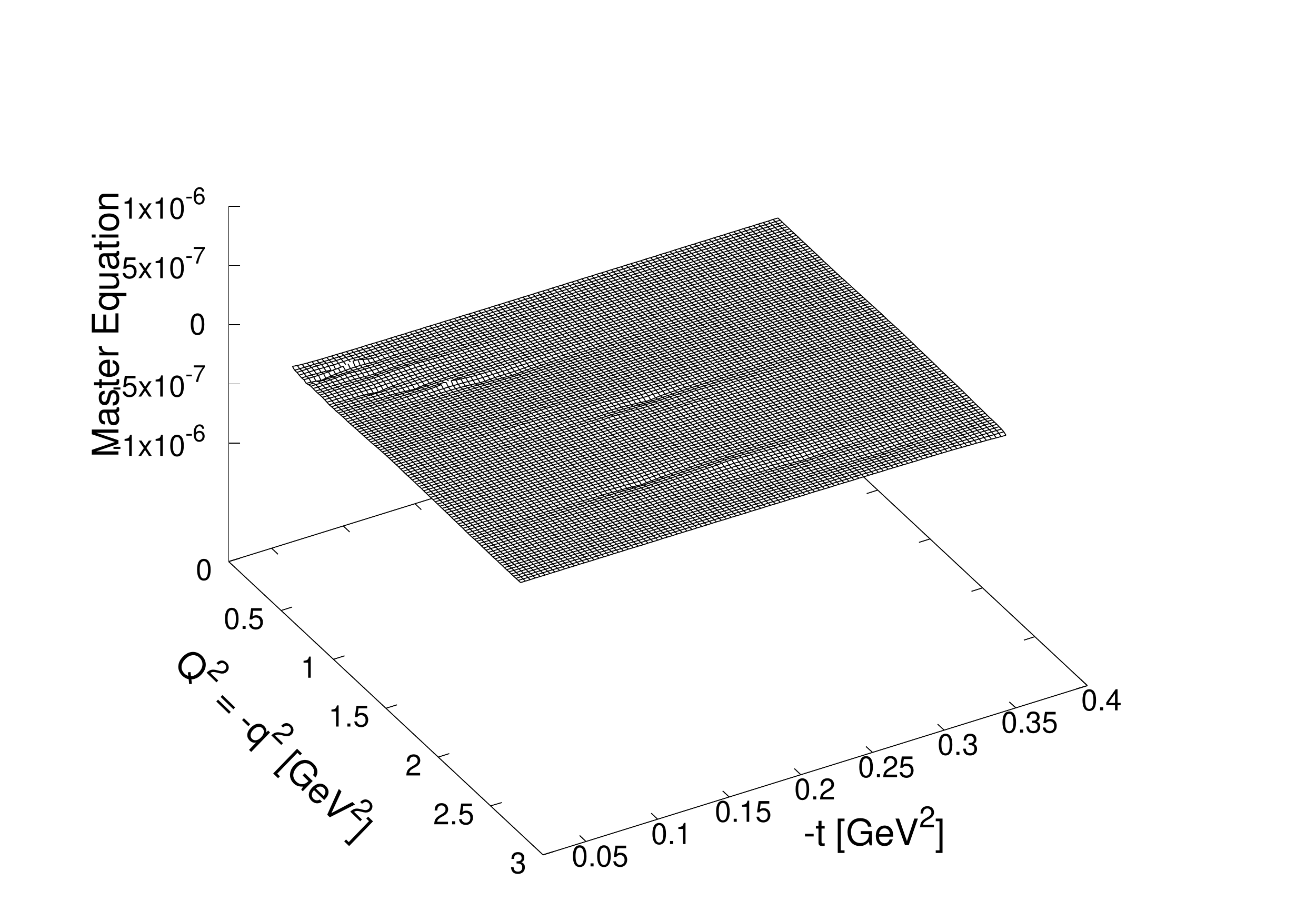}
\caption{\label{fig7} The 3D plot of the master equation~(\ref{eq7}) built with our experimental extraction  
of the off-shell pion form factors.
  In computing Eq.~(\ref{eq7}), we use the values of  $F^{\rm Exp}_1(Q^2,t)$, $F^{\rm Cov}_1(0,t)$, and $g^{\rm Exp}(Q^2,t)$
  given in Table~\ref{t1}.}
\end{center}
\end{figure}

The experimental data  for $d\sigma_{\rm L}/dt$ given in Table VII of Ref.~\cite{Blok2008}  are used for the
extraction of the off-shell form factor $F^{\rm Exp}_1(Q^2,t)$ using Eqs.~(\ref{ex28})-(\ref{ex30}), with the theory input from our model calculation
presented in the previous section, Sec.~\ref{pionhosffmodel}.
Since there are no experimental data available for $F_1(Q^2=0,t)$, we extract $F^{\rm Exp}_2(Q^2,t)$ [or $g^{\rm Exp}(Q^2,t)$] from the WTI 
using the values of $F^{\rm Cov}_1(Q^2=0,t)$ obtained from the manifestly covariant model,  i.e.
$g^{\rm Exp} (Q^2,t)= [F^{\rm Cov}_1(0,t) - F^{\rm Exp}_1(Q^2,t)]/Q^2$. 
For the comparison of the covariant model with the experimental data, we use $m_q=(0.14\pm 0.02)$ GeV,
checking the sensitivity of our covariant model calculation. The experimentally extracted off-shell form factors $F^{\rm Exp}_1(Q^2,t)$ and $g^{\rm Exp}(Q^2,t)$
and the corresponding results from the covariant model obtained from using $m_q=(0.14\pm 0.02)$ GeV 
are summarized in Table~\ref{t1}, in which $(Q^2, -t)$ values are classified into six different sets in terms of average $\la Q^2\ra$ and the invariant mass $W$ following
Ref.~\cite{Blok2008}. 
To check the consistency of our experimental extraction of the form factors, we computed the master equation using  the values of  
$F^{\rm Exp}_1(Q^2,t)$, $F^{\rm Cov}_1(0,t)$ and $g^{\rm Exp}(Q^2,t)$ given in Table~\ref{t1}. The attained 3D plot of the master equation Eq.~(\ref{eq7})
is shown in Fig.~\ref{fig7}. As we have already used Eq.~(\ref{eq7}) to obtain $g^{\rm Exp}(Q^2,t)$, this may be regarded as an obvious cross-check 
just for the purpose of illustration. 

\begin{figure}
\begin{center}
\includegraphics[width=5.5cm,angle=-90]{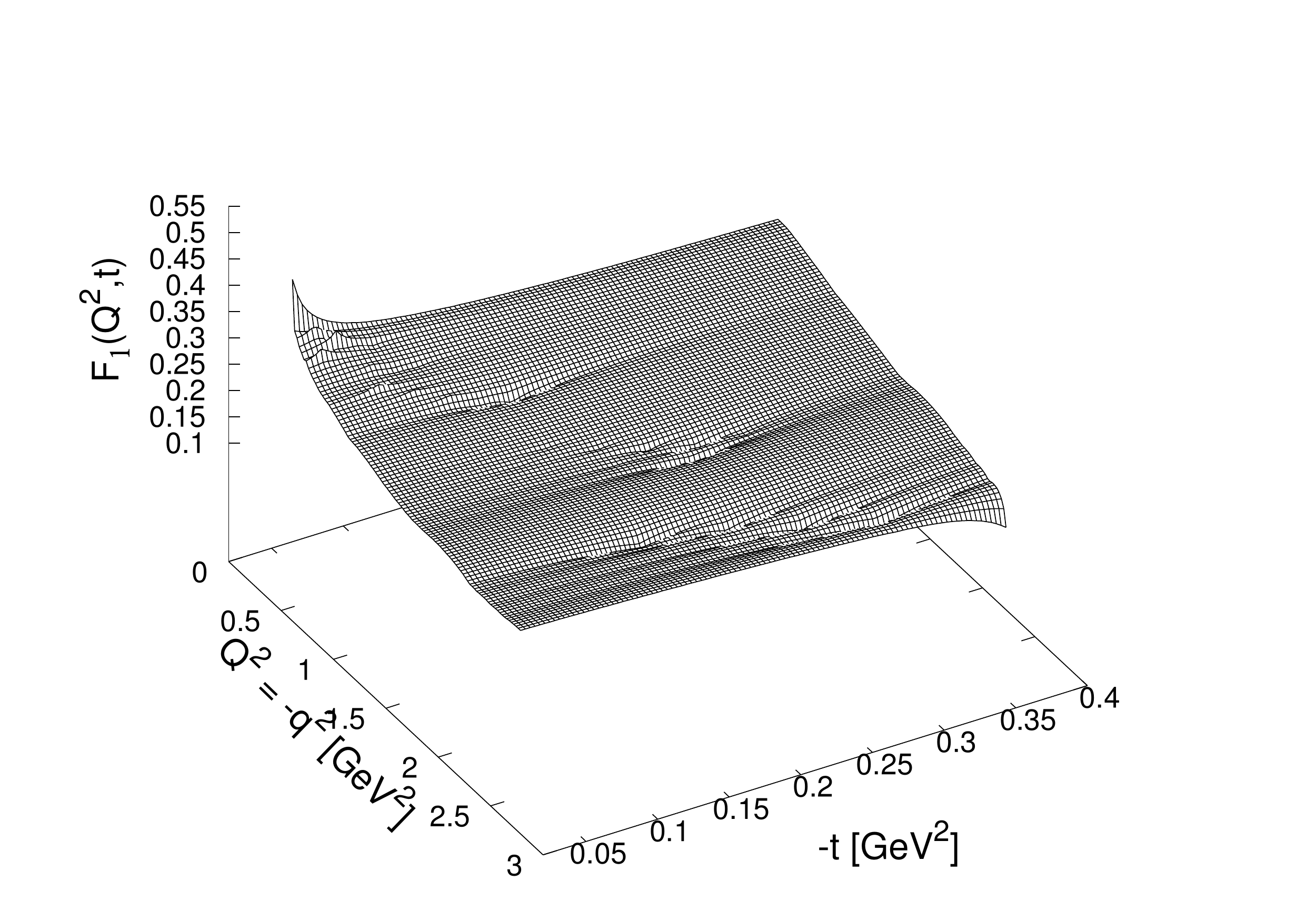}
\includegraphics[width=5.5cm,angle=-90]{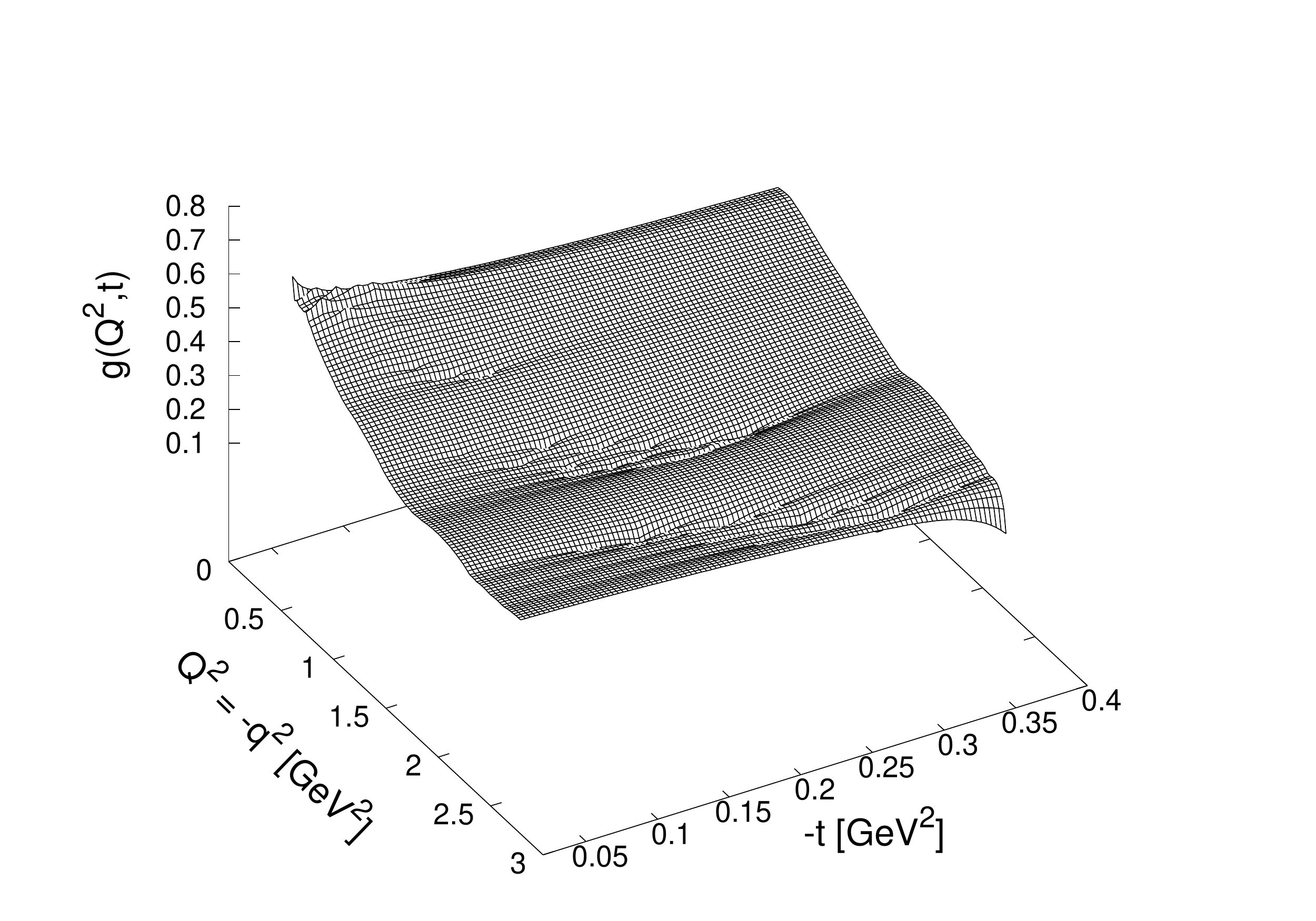}
\caption{\label{fig8} The extracted off-shell pion form factors: $F^{\rm Exp}_1(Q^2, t)$ (top) and   $g^{\rm Exp}(Q^2, t)$ (bottom)
  given in Table \ref{t1}.}
\end{center}
\end{figure}

In Table~\ref{t1}, we note that the $Q^2$ and/or $-t$ evolution of the extracted values of $F^{\rm Exp}_1(Q^2,t)$
is somewhat different from the result of $F_1(Q^2,t)$ due to our covariant analytic model calculation. 
This difference may not be a surprise though, not only due to the simplicity of the covariant analytic model
but also due to the limitation of the Chew-Low extrapolation involving the pion-nucleon form factor 
in crossing the disallowed kinematic region $t>0$ of the electroproduction process. 
While the improvement of the model deserves interest with respect to 
the QCD dynamics of the pion, it suggests the direct extraction of the off-shell pion form factors in lieu of
the extrapolation procedure involving the disallowed kinematic region from the differential cross section of
the electroproduction data.   

The extracted off-shell form factors $F^{\rm Exp}_1(Q^2,t)$ and $g^{\rm Exp}(Q^2,t)$ from the 30 data points in Table~\ref{t1} are plotted 
in Fig.~\ref{fig8} with respect to $Q^2$ and $t$.
The overall momentum dependences of $Q^2 $ and $t$ resemble the results of the covariant analytic model as shown in Fig.~\ref{fig6}. 
While the data seem to exhibit the stronger variation with respect to $Q^2$ and $t$ than the model result 
as also noted in Table~\ref{t1}, the main features captured in the variation appear consistent between 
Figs.~\ref{fig6} and \ref{fig8} from the model calculation and the data extraction, respectively.

\subsection{Comparison of extracted vs model  form factors}\label{results-3}

\begin{figure}
\begin{center}
\includegraphics[height=7cm, width=7cm]{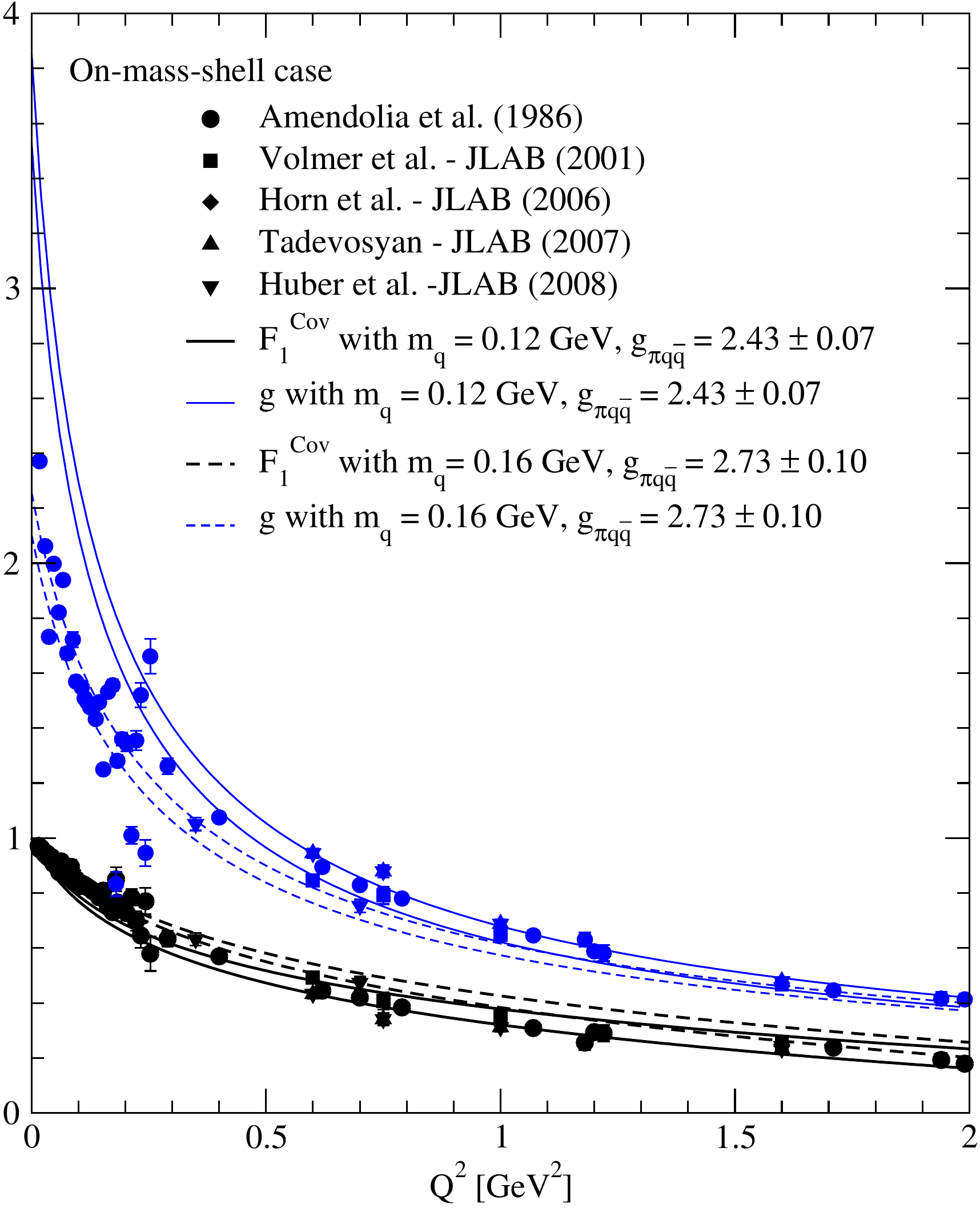}
\caption{\label{fig9} The on-shell pion form factors  $F_1(Q^2, m^2_\pi)$ (black lines) and $g(Q^2, m^2_\pi)$ (blue lines)
for the spacelike momentum transfer region $0\leq Q^2\leq 2$ GeV$^2$ compared with the experimental data for
$F^{\rm Exp}_1$ (black data) and  $g^{\rm Exp}$ (blue data).
The used model parameters are $m_q=(0.12, 0.16)$ GeV using the variation of the couplings 
$g_{\pi q{\bar q}}=(1.32\pm 0.04, 1.11\pm 0.04)(2 m_q /f^{\rm Exp}_\pi)$, respectively, and we show only the
upper and lower limits of $g_{\pi q{\bar q}}$.
}
\end{center}
\end{figure}

The on-shell pion form factors  $F_1(Q^2, m^2_\pi)$ (black lines) and $g(Q^2, m^2_\pi)$ (blue lines) from the covariant model for 
the spacelike region $Q^2>0$ are shown in Fig.~\ref{fig9} and compared with the extracted values of
$F^{\rm Exp}_1(Q^2,t=m^2_\pi)$ (black data) and  $g^{\rm Exp}(Q^2, t=m^2_\pi)=[1 - F^{\rm Exp}_1(Q^2,t=m^2_\pi)]/Q^2$ (blue data).
The model parameters in Fig.~\ref{fig9} are $m_q=(0.12, 0.16)$ GeV using the variation of the couplings 
$g_{\pi q{\bar q}}=(1.32\pm 0.04, 1.11\pm 0.04)(2 m_q /f^{\rm Exp}_\pi)$, respectively.
The solid and dashed lines 
represent the results obtained from  $m_q=0.12$ and 0.16 GeV using 
the upper and lower limits of the corresponding $g_{\pi q{\bar q}}$. 
We should note that while the upper (lower) line of $F_1(Q^2, m^2_\pi)$ corresponds 
to the lower (upper) limit of $g_{\pi q {\bar q}}$, the upper (lower) line of $g(Q^2, m^2_\pi)$ corresponds to the upper (lower) limit of $g_{\pi q {\bar q}}$.
Unlike the form factor $F_2(Q^2,t)$, the form factor $g(Q^2,t)$ does not vanish in the on-shell limit.
We note that the current Particle Data Group~\cite{PDG2018} average $r^{\rm Exp}_\pi =\sqrt{\langle r^2_\pi\rangle}= (0.672\pm 0.008)$ fm for the rms value of 
the pion charge radius corresponds to $g^{\rm Exp}(Q^2=0,m^2_\pi)=(1.953\pm 0.023)$ GeV$^{-2}$. 
Although the more realistic model than the present one may be required to predict $g(Q^2, m^2_\pi)$ more accurately, 
we note that the form factor $g(Q^2, m^2_\pi)$ should be regarded as the physical observable in the on-mass-shell limit
on par with the charge form factor $F_1(Q^2, m^2_\pi)$.
In this respect, it is interesting to observe that $g^{\rm Exp}(Q^2, t=m^2_\pi)=[1 - F^{\rm Exp}_1(Q^2,t=m^2_\pi)]/Q^2$
exhibits a rather large fluctuation near $Q^2=0$, which may reflect a correspondingly large uncertainty in 
determining the pion charge radius.

\begin{figure*}
\begin{center}
\includegraphics[height=7cm, width=7cm]{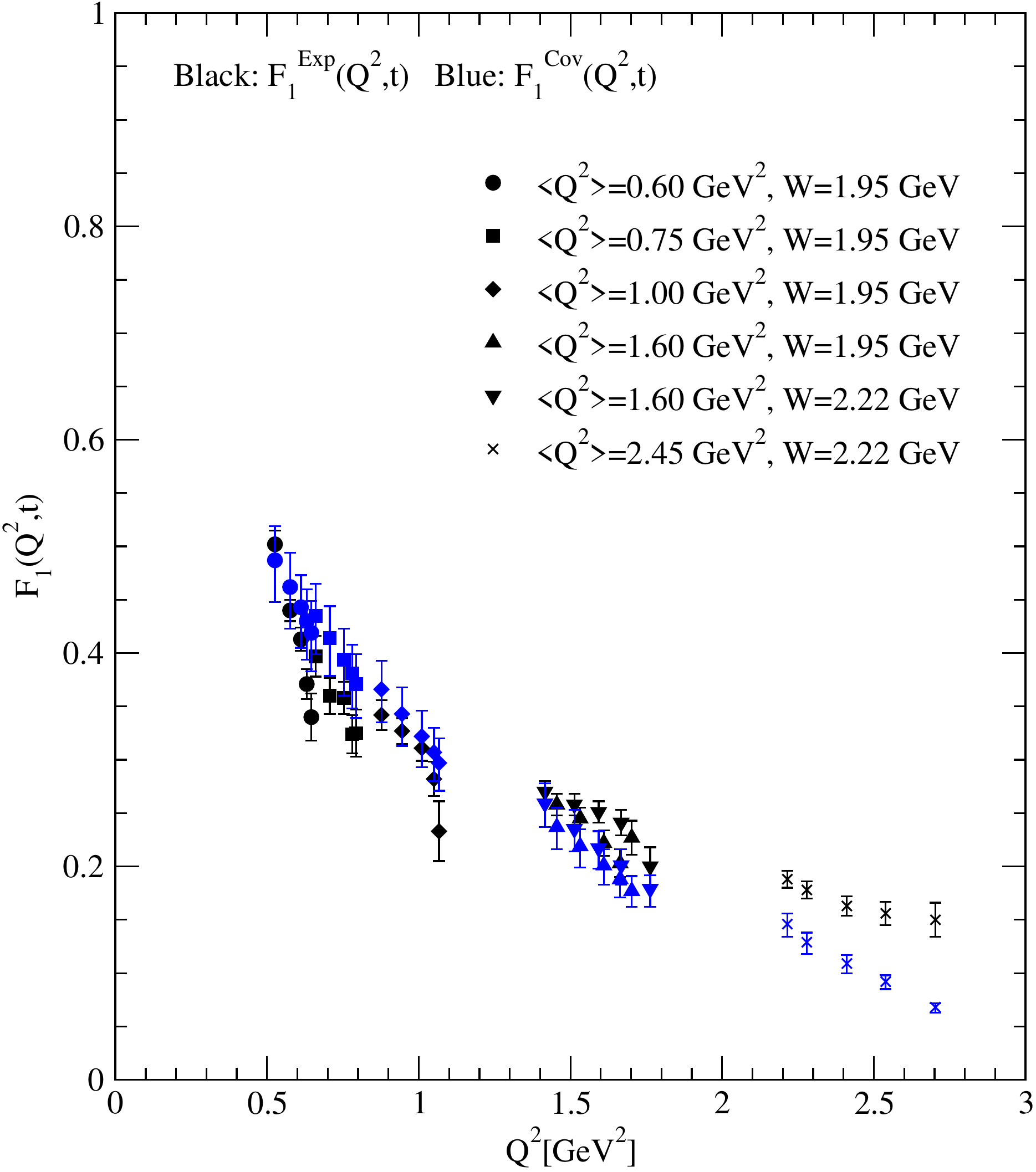}
\includegraphics[height=7cm, width=7cm]{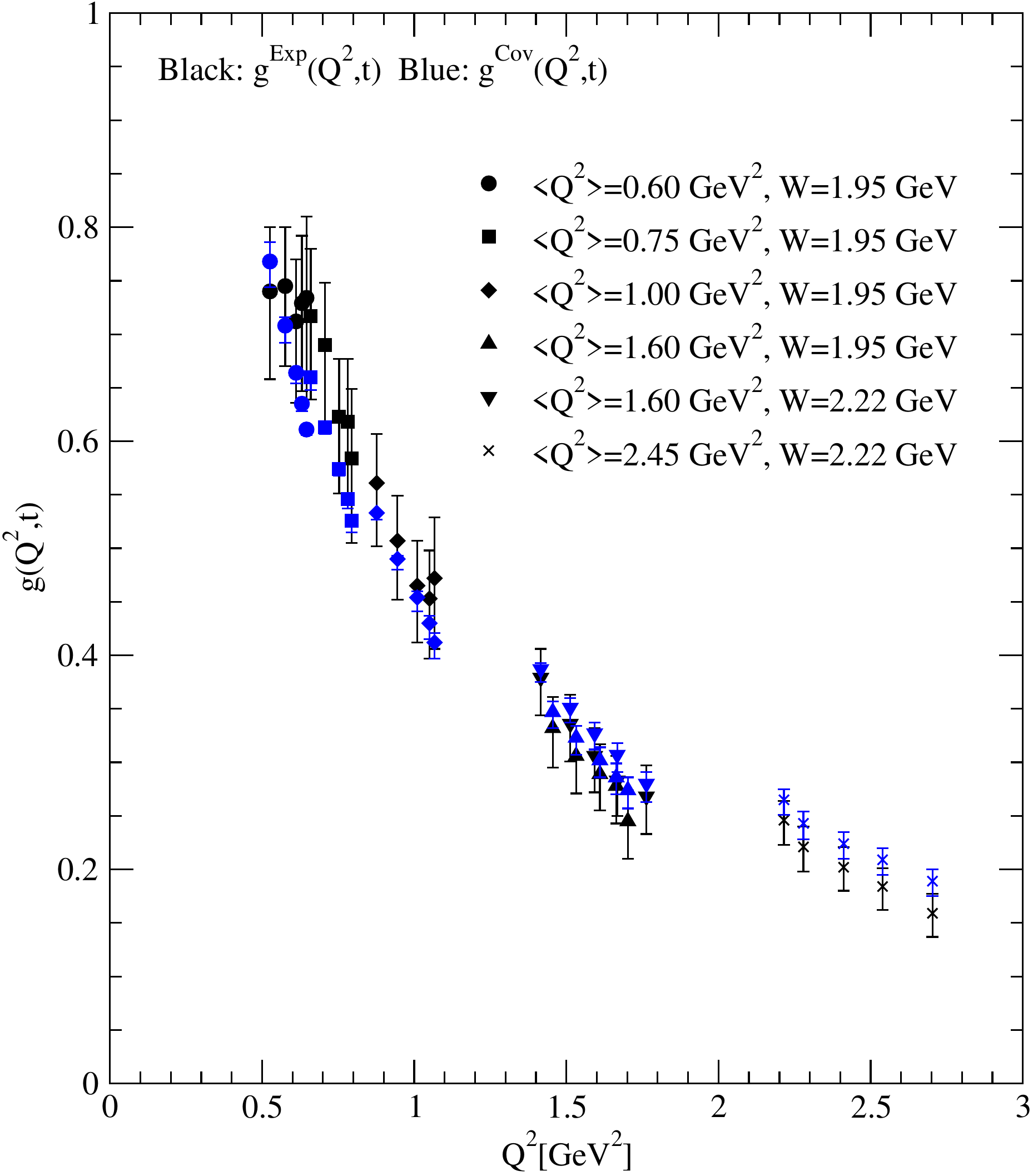}
\includegraphics[height=7cm, width=7cm]{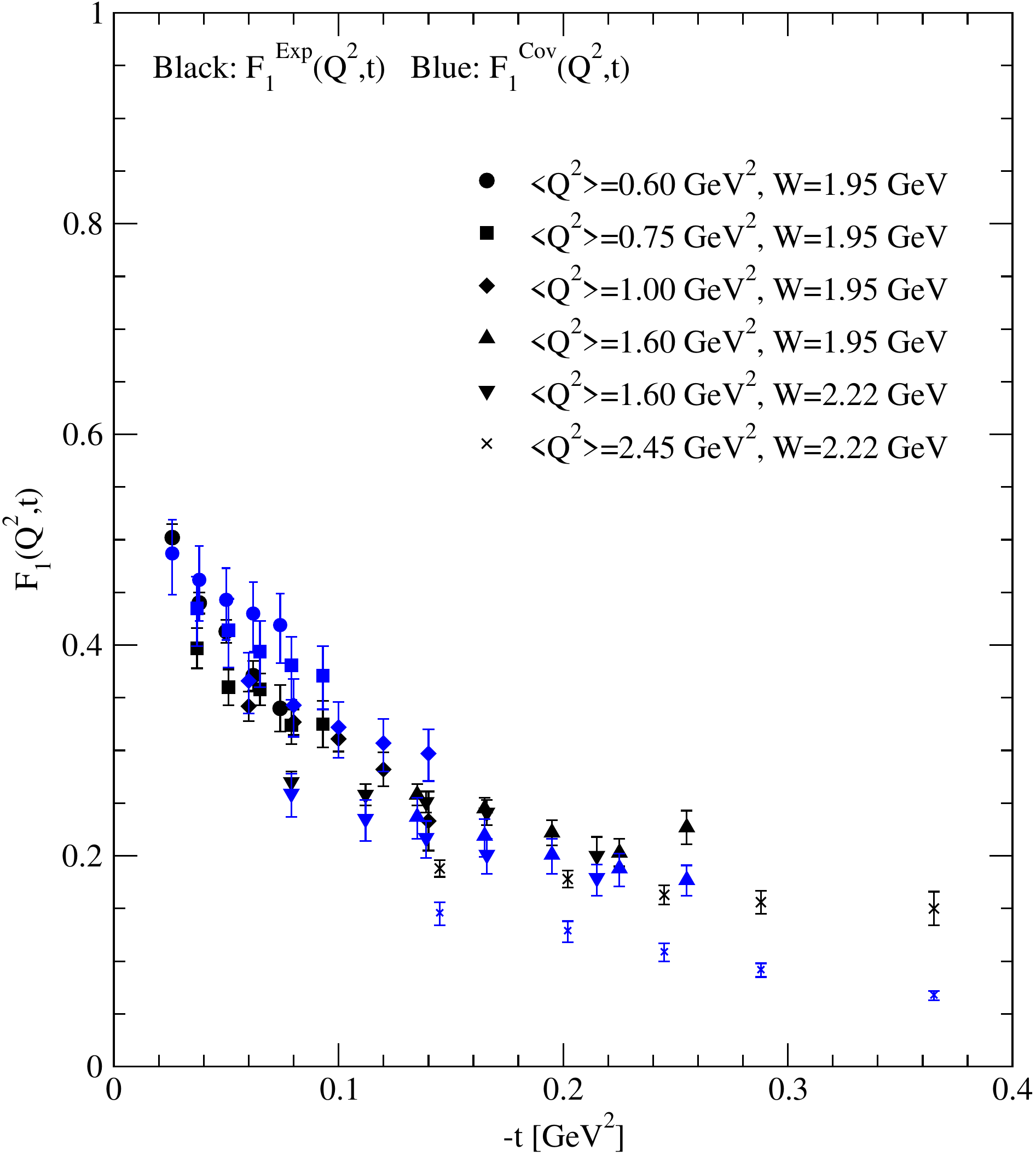}
\includegraphics[height=7cm, width=7cm]{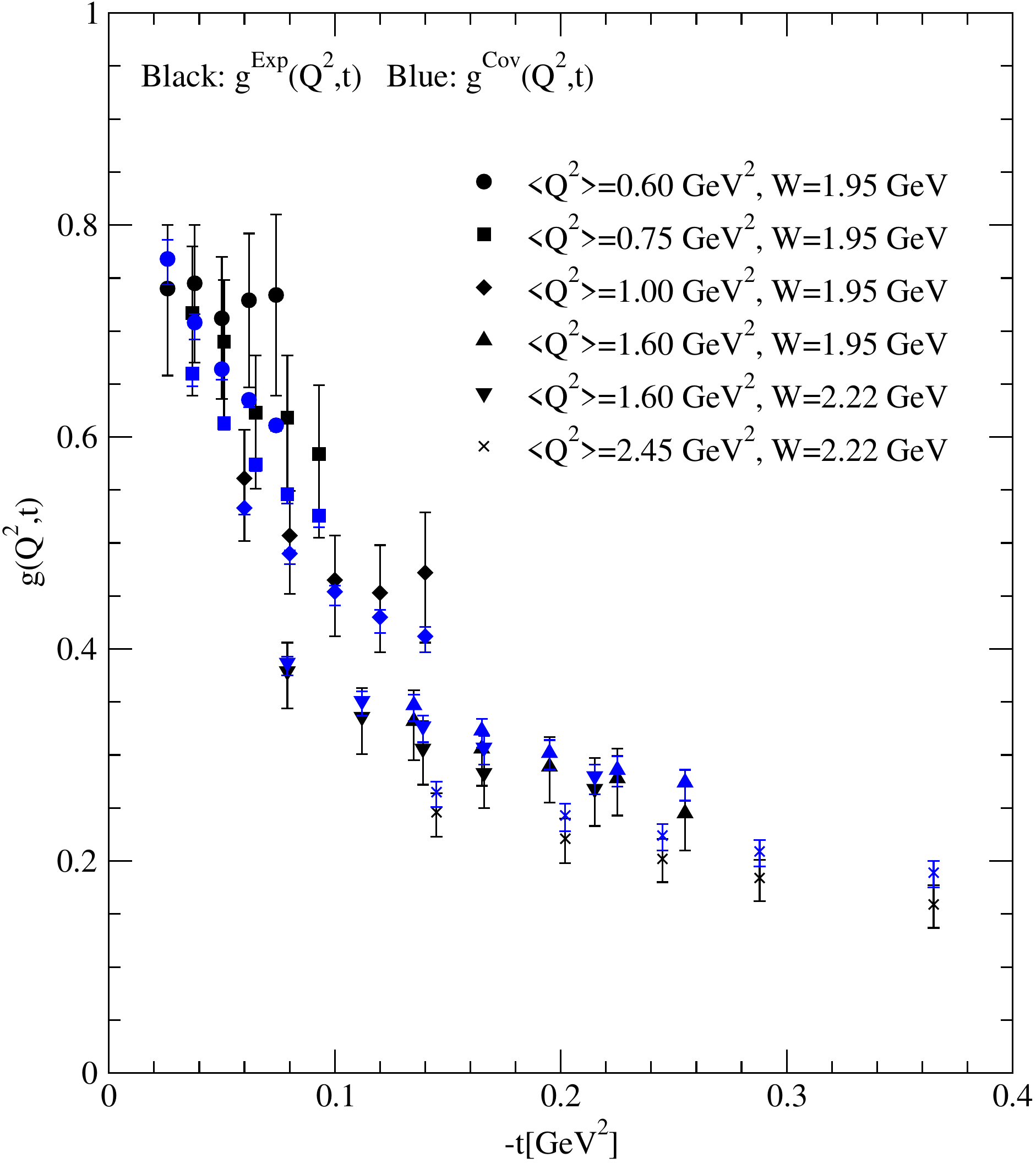}
\caption{\label{fig10} The comparison of the off-shell pion form factors  $F_1(Q^2, t)$  and $g(Q^2, t)$ given in Table~\ref{t1} and those obtained
from the covariant model. The top and bottom panels 
show $Q^2-$ and $t$ dependences of $F_1(Q^2, t)$ (left) and $g(Q^2, t)$ (right), respectively. }
\end{center}
\end{figure*}

The extracted off-shell pion form factors  $F_1(Q^2, t)$  and $g(Q^2, t)$ given in Table~\ref{t1} and 
those obtained from the covariant model are compared  in Fig.~\ref{fig10}. 
The top panel shows the $Q^2$ dependence of $F_1(Q^2, t)$ (left) and $g(Q^2, t)$ (right) collecting 
all the data in Table~\ref{t1} regardless of $t$ values, while bottom panel  
shows the $t$ dependences of $F_1(Q^2, t)$ (left) and $g(Q^2, t)$ (right) collecting 
all the data in Table~\ref{t1} regardless of $Q^2$ values.
The black and blue data represent, respectively, 
the extracted data from the JLab experiment~\cite{Blok2008} and the results of the covariant model obtained from Eqs.~(\ref{off3})~and~(\ref{off3-2})
using the quark mass $m_{u(d)}=0.14\pm 0.02$ GeV.
A rather significant difference in the slope of $Q^2$ evolution between $F^{\rm Exp}_1(Q^2,t)$ and $F^{\rm Cov}_1(Q^2,t)$
in the top left panel of Fig.~\ref{fig10} may be understood from the QCD effect on $F^{\rm Exp}_1(Q^2,t)$ from the gluon exchange 
between the quark and antiquark that gets important as $Q^2$ gets larger, while the solvable model result $F^{\rm Cov}_1(Q^2,t)$ 
does not accommodate this perturbative QCD feature. It is interesting to see, however, that the newly introduced form factor $g(Q^2,t)$ defined by Eq.(\ref{eq8}) appears 
independent of this feature. The model-independent experimental extraction of $F_1(0,t)$ appears indispensable to make the more accurate 
assessment on the $g(Q^2,t)$ behavior without involving any model dependence. 

\section{Conclusions}\label{summary}

In this work, we investigated
the pion electromagnetic half-off-shell form factors $F_1(Q^2,t)$ and
$F_2(Q^2,t)$ using the manifestly covariant fermion field theory
model. 

In this simple covariant model, since our result for 
$F_1(Q^2, t=m^2_\pi)$ includes a UV divergence $1/\epsilon$ term of the dimensional regularization, 
we fixed the normalization of the electric charge via the subtractive renormalization, 
i.e. $F^{\rm ren}_1(Q^2,t) =1+ [F_1(Q^2,t) - F_1(0, m^2_\pi) ]$, as the loop correction
to the charge form factor must vanish at $Q^2=0$.  We used  $F^{\rm ren}_1(Q^2,t)$ as our
off-shell form factor $F_1(Q^2, t)$ throughout the analysis. Our result for
$F_2(Q^2,t)$ is, however, free from the UV divergence. 
In our covariant model with the constituent quark mass $m_q$ as a free parameter, we note that the $\pi q{\bar q}$ coupling
constant $g_{\pi q{\bar q}}$ is related with the pion decay constant $f_{\pi}$ together with the constiuent
quark mass via $g_{\pi q{\bar q}}\approx 2m_q/f_\pi$. In our numerical calculation, however, we used $g_{\pi q{\bar q}}$
as a free parameter to find the best fit for the form factors compared to the experimental data. It turns out that our 
best fits for the constituent quark mass ranges $0.12\leq m_q\leq 0.16$ GeV and the corresponding coupling
$g_{\pi q{\bar q}}$ are consistent with  the values of $2m_q/f_\pi$ within 15$\%$ errors.

We also note that the ratio of $F_2(Q^2,t)$ to $t-m^2_\pi$ is
nonzero in the limit of $t\to m^2_\pi$ while $F_2(Q^2,t)$ goes to zero as $t\to m^2_\pi$.
This led us to define the new form factor $g(Q^2,t)=F_2(Q^2,t)/(t-m^2_\pi)$, 
which should be 
measurable even in the on-mass-shell limit on par with the usual charge form factor $F_1(Q^2, m^2_\pi)$.   
In particular, we obtain the sum rule  
given by Eq.~(\ref{eq7}) which relates
$g(Q^2,t)$ to $F_1(Q^2,t)$ and note that the value of $g(Q^2=0,t=m^2_\pi)$ corresponds to the charge radius of a pion.

According to Eq.~(\ref{eq7}), however, one needs the information of $F_1(0,t)$ to determine $g(Q^2,t)$, while 
no data of $F_1(Q^2,t)$ exist at $Q^2=0$ for $t<0$.  
 In this work, we used a simple covariant model to provide
at least a clear example of demonstration for the simultaneous extraction of 
both $F_1(Q^2,t)$ and $g(Q^2,t)$ (or $F_2(Q^2,t)$). 
In our numerical calculations, we show the 3D plots of $F_{1(2)}(Q^2,t)$ and $g(Q^2,t)$ 
in terms of $(Q^2, t)$ values as shown in Figs.~\ref{fig5} and \ref{fig6}. 

Our extracted values of the pion form factors  obtained 
from the experimental cross section for $d\sigma_{\rm L}/dt$ given in Table VII of Ref.~\cite{Blok2008} and  the results
obtained from the solvable model with $m_q=0.14\pm0.02$ GeV are summarized in Table \ref{t1}.
The extracted off-shell form factors $F^{\rm Exp}_1(Q^2,t)$ and $g^{\rm Exp}(Q^2,t)$ from the 30 data points in Table~\ref{t1} are plotted 
in Fig.~\ref{fig8} with respect to $Q^2$ and $t$.
The main features captured in the variation appear consistent between 
Figs.~\ref{fig6} and \ref{fig8} from the model calculation and the data extraction, respectively.

However, the comparison of the extracted  values of the form factors with covariant model results 
indicates that the evolution in $Q^2$ and/or $t$ are not in full agreement between the extracted vs. model form factors.
On the one hand, this is not unexpected as the internal QCD dynamics of the pion
probed by the electroproduction data should not be restricted only to its valence content,  
while the present model for the pion coupling to the quark and antiquark is just of a pointlike form.
A rather significant difference in the slope of $Q^2$ evolution between $F^{\rm Exp}_1(Q^2,t)$ and $F^{\rm Cov}_1(Q^2,t)$
in the top left panel of Fig.~\ref{fig10} may be an indication of lacking the QCD effect
from the gluon exchange between quark and antiquark that gets important as $Q^2$ gets larger.
The QCD nonperturbative dynamics for the self-energies of quarks and gluons, and the vertices of
pion-quark, photon-quark, etc., deserves further study exploring the 3D imaging of the off-shell form factors. 
On the other hand, the analysis of the electroproduction data by the Chew-Low method demands 
the pion-nucleon form factor as input, which indeed is a simplification and works only close to the pion pole. 
Such a limitation may be also reflected in our extraction of the form factors from
the data, which in part corroborates the difference between the extracted vs model form factors. 

Nevertheless, the overall representation of the trends of the extracted form factors in the $(Q^2,t)$ plane 
by the present constituent model indicates that our analysis goes beyond its obvious limitations. 
It encourages more in-depth theoretical and experimental efforts to reveal the 3D imaging of the off-shell pion form factors.

\acknowledgments
This work was supported in part by the U.S. Department of Energy under
Grant No. DE-FG02-03ER41260 (C.J.);  by the National Research Foundation of Korea (NRF)
under Grant No. NRF-2017R1D1A1B03033129 (H.M.C.); by the project INCT-FNA
Proc. No. 464898/2014-5, 
by CAPES - Finance Code 001, by Conselho Nacional de Desenvolvimento Cient\'ifico e Tecnol\'ogico (CNPq) 
under Grants No. 308025/2015-6 (J.P.B.C.M.), No. 308486/2015-3 (T.F.), and No.
PVE 401322/2014-9 (C.J.); by Funda\c{c}\~ao de Amparo \`a Pesquisa do Estado de S\~ao Paulo (FAPESP) 
under the thematic Projects No. 2013/26258-4 and No. 2017/05660-0; and by regular project 2019/02923-5 (J.P.B.C.M.).
C.J. acknowledges the support from Asia Pacific Center for Theoretical Physics while this work is completed.
This research also used the resources of the National Energy Research Scientific Computing Center (NERSC), 
which is supported by the Office of Science of the U.S. DOE under Contract No. DE-AC02-05CH11231. 

\appendix*
\section{Explicit Calculation of Eqs.~(\ref{off3}) and~ (\ref{off3-2})}
\label{loop-calculation}	

Using the  Feynman parametrization for the three propagators,
we obtain 
\be\label{Con4}
\frac{1}{N_k N_{k+q} N_{p-k}}
=  \int^1_0 \; dx \int^x_0\; dy  \frac{2!}{[ (k+E)^2 - C]^3},
\ee
where
$E = (x-y) q -y p$,
$C = (x-y) (x-y-1) q^2 - y (1-y) t 
- 2y (x-y) q\cdot p + m^2_q$,
and $q\cdot p=(m^2_\pi + Q^2 - t)/2$.

After combining  Eqs.~(\ref{Con1}), ~(\ref{Con2}), and~(\ref{Con4}) and shifting the 4-momentum variable of integration
as $k'=k+E$, we obtain the trace term as
\be\label{TR2}
S^\mu = - 4 ( C^\mu_1 k'^2 + C^\mu_2),
\ee
where
\bea\label{C1C2}
C^\mu_1 &=& \frac{1}{2} [ (1+ 3y) p^\mu + (2 + 3y - 3x) q^\mu],
\nonumber\\
C^\mu_2 &=& p^\mu [ (1+y) (E^2 - m^2_q) -E\cdot q + 2y E\cdot q 
 - y  q\cdot p]
\nonumber\\
&+& q^\mu [ (1-x+y) (E^2 - m^2_q) + (1-2x + 2y) E\cdot p 
\nonumber\\
&&\hspace{3.5cm} +\; (x-y) q\cdot p].
\eea
Using the dimensional regularization
in $d(=4-2\epsilon)$ dimensions, we obtain  the two form factors $F_1(Q^2,t)$ and $F_2(Q^2,t)$ 
from the definition of $\Gamma^\mu = (p'+ p)^\mu F_1(Q^2, t) + q^\mu F_2 (Q^2, t)$ as
given by Eqs.(\ref{off3}) and (\ref{off3-2}), respectively.


\begin{thebibliography}{99}

\bibitem{Dally1} E. B. Dally, D. J. Drickey, J. M. Hauptman, C. F. May, D. H. Stork, J. A. Poirier {\em et al.}, Phys. Rev. D {\bf 24}, 1718 (1981). 

\bibitem{Dally2}  E. B. Dally, J. M. Hauptman, J. Kubic, D. H. Stork, A. B. Watson, Z. Guzik {\em et al.}, Phys. Rev. Lett. {\bf 48}, 375 (1982). 

\bibitem{Amen1} S. R. Amendolia {\em et al.},  Nucl. Phys. {\bf B277}, 168 (1986). 

\bibitem{Amen2} S. R. Amendolia {\em et al.}, Phys. Lett. B {\bf 146}, 116 (1984).

\bibitem{Marco} M. Carmignotto, Ph.D. Dissertation,  The Catholic University of America, 2017.

\bibitem{Sull} J. D. Sullivan, Phys. Rev. D {\bf 5}, 1732 (1972). 

\bibitem{Blok2008} H. P. Blok {\em et al.},  Phys. Rev. C {\bf 78}, 045202 (2008). 

\bibitem{Huber2008} G. M. Huber {\em et al.},Phys. Rev. C {\bf 78}, 045203 (2008). 

\bibitem{JLab3} T. Horn {\em et al.}, Phys. Rev. C {\bf 78}, 058201 (2008). 

\bibitem{JLab4} T. Horn {\em et al.}, Phys. Rev.  Lett. {\bf 97}, 192001 (2006); 
J. Volmer {\em et al.},  Phys. Rev. Lett. {\bf 86}, 1713 (2001); V. Tadevosyan {\em et al.},  Phys. Rev. C {\bf 75}, 055205 (2007).

\bibitem{JLab5} T. Horn and C. D. Roberts,  J. Phys. G {\bf 43}, 073001 (2016).

\bibitem{Rudy} T. E. Rudy, H. W. Fearing, and S. Scherer,  Phys. Rev. C {\bf 50}, 447 (1994). 

\bibitem{Weiss} C. Weiss,  Phys. Lett. B {\bf 333}, 7 (1994).

\bibitem{Craig} S.-X. Qin, C. Chen, C. Mezrag, and C. D. Roberts, Phys. Rev. C {\bf 97}, 015203 (2018). 

\bibitem{CraigK} F. Gao, L. Chang, Y.-X. Liu, C. D. Roberts, and P. C. Tandy, Phys. Rev. D {\bf 96}, 034024 (2017). 

\bibitem{N1} A. M. Bincer, Phys. Rev. {\bf 118}, 855 (1960).

\bibitem{N2}  H. W. L. Naus and J. H. Koch, Phys. Rev. C {\bf 36}, 2459 (1987). 

\bibitem{N3} P. C. Tiemeijer and J. A. Tjon, Phys. Rev. C {\bf 42}, 599 (1990).

\bibitem{Ward} J. C. Ward, Phys. Rev. {\bf 78}, 182 (1950).

\bibitem{Taka} Y. Takahashi, Nuovo Cimento {\bf 6}, 371 (1957).

\bibitem{N4} X. Song, J. P. Chen, and J. S. McCarthy,  Z. Phys. A  {\bf 341}, 275 (1992).

\bibitem{N5} J. W. Bos and J. H. Koch,  Nucl. Phys. {\bf A563}, 539 (1993).


\bibitem{Nis} K. Nishijima, Phys. Rev. {\bf 122}, 298 (1961).

\bibitem{Bar} G. Barton, {\em Introduction to Dispersion Techniques in Field Theory} (Benjamin, New York, 1965).

\bibitem{Naus1998} 
  H.~W.~L.~Naus, J.~P.~B.~C.~de Melo and T.~Frederico,
  Few-Body Syst.\  {\bf 24}, 99 (1998).

\bibitem{PedlarPRL05} T. K. Pedlar  {\em et al.}, Phys. Rev. Lett. {\bf 95}, 261803 (2005).

\bibitem{SethPRL13} K. K. Seth, S. Dobbs, Z. Metreveli, A. Tomaradze, T. Xiao, and G. Bonvicini,  Phys. Rev. Lett. {\bf 110}, 022002 (2013).

\bibitem{VGL}  M. Vanderhaeghen, M. Guidal, and J. M. Laget, Phys. Rev. C {\bf 57}, 1454 (1998).

 \bibitem{PDG2018}  M. Tanabashi {\em et al.} (Particle Data Group), Phys. Rev. D {\bf 98}, 030001 (2018). 
\end{thebibliography}
\end{document}